\documentclass[preprint]{aastex}

\received{}
\revised{}
\accepted{}
\ccc{}
\cpright{}{}

\slugcomment{AJ, in press, Jan 2003}
\shorttitle{Abundance Analysis of M5}
\shortauthors{Ram\'{\i}rez \etal}

\newcommand{\kms}{km~s$^{-1}$}
\newcommand{\subsun}{\mbox{$_{\odot}$}}
\newcommand{\etal}{{\it et al.\/}}
\newcommand{\teff}{$T_{eff}$}
\newcommand{\grav}{log($g$)}
\newcommand{\mtv}{$\xi$}
\newcommand{\ew}{$W_{\lambda}$}
\newcommand{\fe}{[Fe/H]}

\begin{document}

\title{Abundances in Stars from the Red Giant Branch Tip to Near the 
Main Sequence Turn Off in M5.
\altaffilmark{1}}

\author{Solange V. Ram\'{\i}rez \altaffilmark{2,3} and  
Judith G. Cohen\altaffilmark{2}.}

\altaffiltext{1}{Based on observations obtained at the
W.M. Keck Observatory, which is operated jointly by the California 
Institute of Technology, the University of California, and the
National Aeronautics and Space Administration.}

\altaffiltext{2}{Palomar Observatory, Mail Stop 105-24,
California Institute of Technology.}

\altaffiltext{3}{Current Address: SIRTF Science Center, Mail Stop 220-06,
California Institute of Technology.}

\begin{abstract}

We present the iron abundance and abundance ratios for 18 elements 
with respect to Fe in a sample of stars with a wide range in 
luminosity from luminous giants to stars near the turnoff in the 
globular cluster M5.
The analyzed spectra, obtained with HIRES at the Keck Observatory,
are of high dispersion (R=$\lambda / \Delta \lambda$=35,000).
We find that the neutron capture, the iron peak and the 
$\alpha-$element abundance ratios show no trend with \teff, 
and low scatter around the mean between the
top of the RGB and near the main sequence turnoff,
suggesting that 
at this metallicity, non-LTE effects are not important over the range
of stellar parameters spanned by our sample.
To within the precision
of the measurements ($\pm\sim0.1$ dex), 
gravitationally induced heavy element diffusion
does not appear to be present among the stars near the main sequence
turnoff studied here.  Our work and other recent studies suggest that
heavy element diffusion is inhibited in the surface layers of metal poor
stars.

Differences in the Na abundance from star to star which extend  
to the main sequence turnoff are detected in our sample in M5.
The anti-correlation between O and Na abundances, observed in other 
metal poor globular clusters, is not detected in our sample,
but it may be hidden among stars with only upper limits for their O abundances.
As we found in M71, there is a hint of star-to-star variation in the 
Zr abundance.

Overall the abundance ratios of M5 appear very similar to those
of M71, with the possible exception of the neutron capture element Ba,
where we argue that the apparent difference may be due to difficulties
in the analysis.
As in M71, the $\alpha-$elements Mg, Ca, Si and Ti are overabundant 
relative to Fe.

The results of our abundance analysis of 25 stars in M5 provide
further evidence of abundance variations among specific light elements
at unexpectedly low luminosities,
which cannot be explained by our current understanding of stellar evolution.

\end{abstract}


\section{INTRODUCTION}

Abundance determinations of stars in Galactic globular clusters can provide 
valuable information about important astrophysical processes such as
stellar evolution, stellar structure, Galactic chemical evolution and
the formation of the Milky Way. Surface stellar abundances of C, N, O,
and often Na, Mg, and Al are found to be variable among red giants within 
a globular cluster. 
The physical process responsible for these star-to-star element variations 
is still uncertain (see the reviews of Kraft 1994 and
Pinsonneault 1997, as well as Cohen \etal\ 2001 and Ventura \etal\ 2001).  

In order to study the origin of the star-to-star abundance variations,
we started a program to determine chemical abundances of the nearer
galactic globular cluster stars. 
In our first series of papers, we studied 
a sample of stars in M71, the nearest globular 
cluster reachable from the northern hemisphere \citep{coh01,ram01,ram02}.
Our sample of 25 M71 stars includes stars over a large range in luminosity,
in order to study in a consistent manner red giants, horizontal
branch stars, and stars at the main sequence turnoff.
We measured the iron abundance and the abundance ratios for 23 elements 
with respect to Fe in our sample of M71 stars, using high dispersion 
(R=$\lambda / \Delta \lambda$=35,000) optical spectra obtained with 
HIRES at the Keck Observatory.
We found that the \fe\ abundances\footnote{The 
standard nomenclature is adopted; the abundance of
element $X$ is given by $\epsilon(X) = N(X)/N(H)$ on a scale where
$N(H) = 10^{12}$ H atoms.  Then
[X/H] = log$_{10}$[N(X)/N(H)] $-$ log$_{10}$[N(X)/N(H)]\subsun, and similarly
for [X/Fe].}
from both Fe I (\fe\ = $-0.71 \pm 0.08$)
and Fe II (\fe\ = $-0.84 \pm 0.12$) lines agree with each other and with
earlier determinations and that the \fe\ obtained from Fe I and Fe II lines 
is constant within the rather small uncertainties over the full
range in effective temperature (\teff) and luminosity \citep{ram01}.
In \citet{ram02}, we found that the neutron capture, the iron peak and the
$\alpha-$element abundance ratios
show no trend with \teff, and low scatter around the mean between the
top of the RGB and near the main sequence turnoff.
We detected an anti-correlation between O and Na abundances in our sample
of members of M71 which extends to the main sequence.
We also observed a statistically significant correlation between Al and Na 
abundances extending to $M_V = +1.8$, fainter than the luminosity of the 
RGB bump in M5.

In the present paper, we have studied a sample of 25 stars in the globular
cluster M5, again covering a wide range in luminosity. M5
is the nearest intermediate metallicity globular cluster accessible
from a northern hemisphere site.  
We adopt current values
from the on-line database of \cite{har96}
for its apparent distance modulus at $V$ of 14.31 mag
with a reddening of E(B--V) = 0.03 mag.  
Recent CCD photometric studies of this cluster,
focusing primarily on its age,
are given by \cite{joh98} and \cite{ste99}.
\cite{san96} discuss the predominantly blue
horizontal branch of M5.  
Previous high dispersion abundance studies, the most extensive of which is
\citet{iva01}, include only luminous giant and asymptotic giant
branch stars. A study of C and N variations among a large sample
of M5 stars at the base of the red giant branch is given in \cite{coh02}.

Important differences with M71 include the lower metallicity of M5,
its much lower reddening, its location further from the
galactic plane and its larger
radial velocity.   M5 is more distant than M71, 
but this is largely compensated by the
higher reddening of M71.  Hence
the apparent brightness at $V$ of stars at
a given evolutionary stage is roughly equal in the two clusters.

\section{OBSERVATIONS}

To the maximum extent possible,
the observing strategy, the atomic data and the analysis procedures
used here are identical to those developed in our earlier papers on M71
\citep{coh01,ram01,ram02}.

\subsection{The Stellar Sample}

Stars were chosen to span the range from the tip of the red giant branch
to the main sequence turnoff of M5.   This cluster lies considerably
further from the galactic plane at $b = -23^{\circ}$ than does M71,
hence field star contamination is a much less serious problem for M5.
The photometric database of \cite{ste98} and \cite{ste00}, which is
described in considerable detail in \cite{coh02}, 
was used to verify that the  
selected stars lie on the cluster locus in various color-magnitude
diagrams.  For the more luminous
stars, we required the assignment of a high probability 
of membership by \cite{cud79} in
his proper motion survey of this globular cluster. Only reasonably isolated stars were selected.  

Throughout this paper, 
the star names are from \citet{arp62} for the brightest stars, 
from \cite{buo81} or, for the fainter stars previously not catalogued,
are assigned based on the object's 
J2000 coordinates, so that a star with RA, Dec of 
15 rm rs.s +2 dm dd is identified in this paper with
the name Grmrss\_dmdd.

\subsection{Data  Acquisition and Reduction}

All spectra were obtained with HIRES \citep{vog94} at 
the Keck Observatory.  A maximum
slit length of 14 arc sec can be used with our instrumental configuration
without orders overlapping.  Since an image rotator for HIRES is available
(built under the leadership of David Tytler), if we can find pairs of program stars
with separations less than 8 arcsec, they can be observed together on
a single exposure.  Pairs were pre-selected to contain two members 
of the M5 sample.
All stars were observed in pairs, except for the very bright star
IV-59, for which there was no suitable nearby star with which to form
a pair.  One set of three suitable stars was found which fit within the 
maximum allowed slit length. 

The desired minimum SNR was 75 over a 4 pixel resolution element for
a wavelength near the center of echelle order 58 ($\sim$6150 \AA).
This is calculated strictly from the counts in the object spectrum, and 
excludes noise from cosmic ray hits, sky subtraction, flattening problems, etc. 
Since the nights were dark, sky subtraction is not an issue except at
the specific wavelengths corresponding to strong night sky emission lines, 
such as the Na D doublet.  This SNR goal was
achieved, at considerable cost in observing time, for most of the stars,
although we did not meet this goal for the six main sequence turnoff
region stars included in our sample.  Note that for a fixed 
SNR in the continuum, for a star of a given luminosity,
the lower metallicity
of M5 leads to weaker absorption lines, making it difficult to 
maintain the desired precision of the analysis.

\begin{figure}
\epsscale{0.7}
% Comment out the following line to embed the PS figure into the manuscript
\plotone{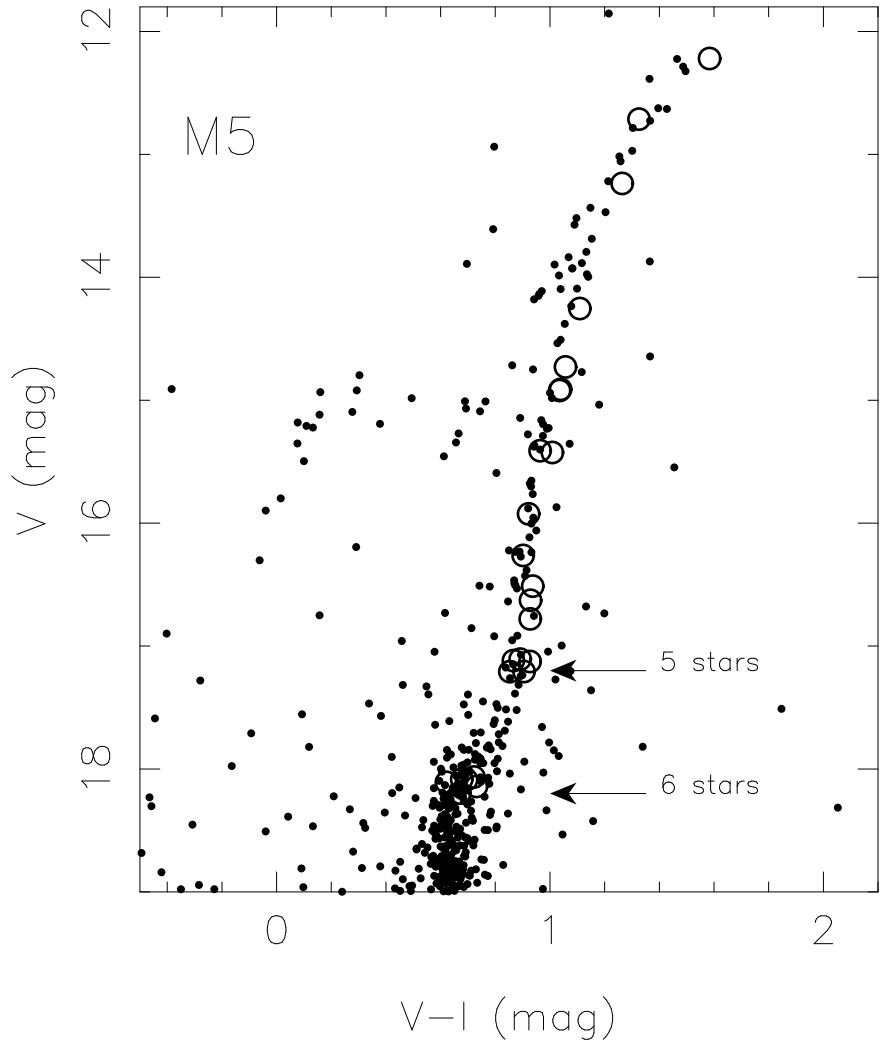}
\caption[m05_hr.ps]{HR diagram of M5. The photometry comes from \citet{ste98,
ste00}. The open circles mark the position of the stars in our M5 sample
observed with HIRES (excluding two BHB stars).
\label{m05_hr}}
\end{figure}

Approximate measurements of the radial velocity were made on line,
and if a star was determined to be a non-member, the observations were 
terminated.  Very few non-members turned up in this way.
If the probable non-member was the second component in a pair, an attempt
was made to switch to another position angle to pick up a different 
second star, when a possible candidate that was bright enough 
was available within the limits of the 8 arcsec maximum separation.
Through creative use of close pairs, a sample of 27 members of M5
were observed
with HIRES, two of which are blue horizontal branch stars in M5;
we subsequently ignore them.  Figure~\ref{m05_hr} shows the sample
in M5 superposed on a color-magnitude diagram of this globular cluster.

The observations were centered
at about 6500\AA, as were the M71 spectra, with the reddest order
reaching the O triplet.
Because the HIRES detector is undersized, our spectra do not cover the 
full length of each echelle order without gaps in the wavelength coverage.
We wanted to include key lines of critical elements,
specifically the 6300, 6363 [OI] lines, the 7770 O triplet, the
Na doublet at 6154, 6160\AA, and the 6696, 6698\AA\ Al I lines.  
However, it was impossible to create a single instrumental
configuration which included all
the desired spectral features in the wavelength range 6000 to 8000 \AA,
and a single compromise configuration had to be adopted.
In particular, although the 6696, 6698\AA\ Al I doublet is the most useful 
feature of that element in this spectral region, we could not get it
to fit into a single HIRES setting together with the O lines.  

The spectra of the brighter stars in our M5 sample were observed
on two nights in June 2000.  
A 1.15 arcsec slit was used, which provides a spectral
resolution of 34,000.  All long integrations were broken 
up into separate exposures, each 1200 sec long, to optimize cosmic
ray removal.
These data were reduced by Brad Behr using Figaro 
(Shortridge 1993) scripts 
with commands written by McCarthy and Tomaney (McCarthy 1988)
specifically for echelle data reduction.   Observations
continued with the same instrumental configuration
during three nights in May 2001 to cover
the fainter stars in the sample.  This set of HIRES data was reduced
by JGC using a combination of Figaro scripts and
the software package MAKEE\footnote{MAKEE was developed
by T.A. Barlow specifically for reduction of Keck HIRES data.  It is
freely available on the world wide web at the
Keck Observatory home page, http://www2.keck.hawaii.edu:3636/.}.

Table 1 gives details of the HIRES exposures for each star, with the
total exposure time for each object.
The signal to noise ratio per 4 pixel spectral resolution element 
in the continuum at 6150 \AA\ is also given, 
calculated assuming Poisson statistics and ignoring
issues of cosmic ray removal, flattening etc.  The latter become non-negligible
for the very long HIRES integrations necessary 
to reach as faint as possible in M5.
Also listed in the last column
is the radial velocity for each star,
measured from the HIRES spectra as described in \S\ref{radial_velocity}.

\subsection{Radial Velocities \label{radial_velocity}}

Radial velocities were measured from all the M5 spectra using
a list of 36 strong isolated features within the wavelength range of
the HIRES spectra with laboratory wavelengths from the NIST Atomic Spectra 
Database Version 2.0 (NIST Standard Reference Database \#78).
Using an approximate initial $v_r$,
the list of automatically detected lines, restricted to the
strongest detected lines only, generated in the
course of measuring the equivalent widths of the lines 
(see \S\ref{equiv_widths}) in the spectrum of
each of the M5 stars was then searched for
each of these features.  A $v_r$ for each line was determined
from the central wavelength of the best-fit Gaussian, and the
average of these defined the $v_r$ for the star.
Heliocentric corrections appropriate for each exposure were then applied
to the measured $v_r$\footnote{While the MAKEE reduction package
removes this term automatically, heliocentric corrections must be
calculated and 
explicitly removed from echelle spectra analyzed using Figaro scripts.}.
The radial velocities for our sample of stars is listed
in column 6 of Table~\ref{tab_sample}.
Based on their measured radial velocities, all the stars of our
sample are members of M5.
They have a mean $v_r$ of +55.0 \kms, with   $\sigma$ = 4.3 \kms.  
This is in good agreement with the value of \citet{har96} of
$v_r = +52.6$ \kms.  The velocity dispersion we find for M5 is also in
good agreement with the  published value of 4.9 \kms\ for
a large sample of luminous giants from
\cite{rastor91}.

\section{EQUIVALENT WIDTHS  \label{equiv_widths}}

The search for absorption features present in our HIRES data and the
measurement of their equivalent width (\ew) was done automatically with
a FORTRAN code, EWDET, developed for our globular cluster project. 
Details of this code and its features are described in \citet{ram01}.
Because M5 is considerably more metal poor than M71, the determination
of the continuum level was easier, and the equivalent widths
measured automatically should be more reliable.

A list of unblended atomic lines 
with atomic parameters was created by inspection of the 
spectra of M5 stars, as well as the online Solar spectrum taken with 
the FTS at the National Solar Observatory of Wallace, Hinkle \& 
Livingston (1998) and the set of Solar line identifications of 
\citet{moo66}.
The list of lines identified and measured by EWDET is then correlated,
taking the radial velocity into account, 
to the list of suitable unblended lines  
to specifically identify the various atomic lines. 

In \citet{ram01}, we derived $\lambda D-$\ew\ relation (where $D$ is
the central depth of the line)  for the Fe I 
lines of ``the weak line set'' (Fe I lines within 2$\sigma$ of the 
$\lambda D-$\ew\ fit, with \ew\ $<$ 60 m\AA, and errors less than a 
third of the 
\ew) for our analysis of similar spectra in the globular cluster M71.
We used these $\lambda D-$\ew\ relations to determine ``the good line set''
(lines with errors less than a third of the \ew\ and with \ew\
computed from the derived $\lambda D-$\ew\ relations) in the M71 data.
We have used the same approach for the analysis of the atomic spectral lines
present in the spectra of our M5 sample. 
We computed the $\lambda D-$\ew\ fit for each star's Fe I lines, except
for the six faintest stars, where not enough Fe I lines were detected
to do so. For the six faintest stars we used for the reminder of the
analysis the equivalent widths measured automatically by EWDET.
We used the $\lambda D-$\ew\ relations to determine 
``the weak line set'' for the Fe I lines and the 
``the good line set'' for the lines of all the elements, except 
for the O I and Ca I lines, and for the elements that show hyperfine 
structure splitting (Sc II, V I, Mn I, Co I, Cu I, and Ba II). 
The equivalent widths of the O I lines were measured by hand, since 
thermal motions become important at its low atomic weight and the 
$\lambda D-$\ew\ relations derived for Fe I lines may no longer be valid. 
For Ca I lines and the lines of elements that show hyperfine structure 
splitting, we used the equivalent widths measured automatically by EWDET,
but did not force them to fit the Fe I $\lambda D-$\ew\ relationship due 
to the probable different broadening mechanisms. A few
of the Ca I lines were strong enough to be on the damping part of 
the curve of growth, but only in the spectra of the coolest M5 giants.
The \ew\ used in the abundance analysis are listed in Table~\ref{tab_ew} 
(available electronically).

\section{ATOMIC PARAMETERS}

The provenance of the $gf$ values and damping constants we adopt
in our analysis of M5 stars
is discussed below.  In general, the atomic data and the analysis procedures
used here are identical to those developed in our earlier papers on M71
\citep{coh01,ram01,ram02}.  
At the request of the referee, however, we have updated the Solar 
equivalent widths we use to derive the adopted Solar abundances
(see \S\ref{section_sun}).
Only small changes in the resulting abundances are introduced by 
these modifications.

\subsection{Transition Probabilities}

Transition probabilities for the Fe I lines were obtained from several
laboratory experiments, including studies of Fe I absorption
lines produced by iron vapor in a carbon tube furnace
\citep{bla79,bla82a,bla82b,bla86} (Oxford Group), measurement of 
radiative lifetimes of Fe I transitions by laser induced fluorescence 
\citep{obr91,bar91,bar94}, Fe I emission line spectroscopy from a low 
current arc \citep{may74}, and emission lines of Fe I from a shock 
tube \citep{wol71}.
We also considered solar $gf$ values from \citet{the89,the90} when needed.
The Fe I $gf$ values obtained by the different experiments were placed
into a common scale with respect to the results from
\citet{obr91} (see \cite{ram01} for details).
The $gf$ values for our Fe II lines were taken from the solar analysis of
\citet{bla80}, \citet{bie91a}, and from the semiempirical calculations of
\citet{kur93b}. For Fe I and Fe II $gf$ values, we used the same priority 
order for the $gf$ values from different experiments as in \citet{ram01}.

Transition probabilities for the lines of atomic species other than iron
were obtained from the NIST Atomic Spectra Database (NIST Standard
Reference Database \#78, see \citep{wei69,mar88,fuh88,wei96}) when possible.
Nearly 80\% of the lines selected as suitable from the
HIRES spectra have transition probabilities from the NIST database.
For the remaining lines the $gf$ values come from the inverted solar
analysis of \citet{the89,the90},
with the exception of La II and Eu II lines (see \S\ref{section_sun}). 
The solar $gf$ values of Mg I, Ca I, Ti I, Cr I, and Ni I were corrected 
by the factors derived earlier (see Table~3 of Ram\'{\i}rez \& Cohen 2002) 
which are needed to place both sets of transition probabilities onto the 
same scale.  
The correction factors were computed as the mean difference
in log($gf$) between the NIST and solar values for the lines in common.

Six elements show hyperfine structure splitting (Sc II, V I, Mn I, Co I, Cu I,
and Ba II). The corresponding hyperfine structure constants were taken
from \citet{pro00}.

\subsection{Damping Constants}

As in our M71 work, the 
damping constants for all Fe I and Fe II lines were set to twice
that of the Uns\"{o}ld approximation for van der Waals broadening
following \citet{hol91}.
Some of the Na I and Ca I lines are strong enough for damping effects
to be important.
For Na I the interaction constants, $C_{6}$, of the van der Waals broadening 
were taken from the solar analysis of \citet{bau98}. 
\citet{smi81} studied collisional broadening of 17 Ca I lines. 
Comparing their experimental results and the predicted values of $C_{6}$
obtained using the Uns\"{o}ld approximation, we found that the 
experimental $C_{6}$ are comparable to the Uns\"{o}ld $C_{6}$. 
Thus for the Ca I we used the experimental $C_{6}$
from \citet{smi81} when available; if not, we set  $C_{6}$ to be
that of the Uns\"{o}ld approximation.
The empirical values of $C_{6}$ for Mg I from \citet{zha98} are also used.
For the lines of all other ions we set $C_{6}$ to be
twice the Uns\"{o}ld approximation following \citet{hol91}.

\subsection{Solar Abundances \label{section_sun}}

We need to establish the solar abundances corresponding to our 
adopted set of $gf$ values and damping constants.  We measured
our own set of Solar equivalent widths from the online Solar 
spectrum taken with the FTS at the National Solar Observatory of 
Wallace, Hinkle \& Livingston (1998) using our code EWDET.
Solar abundance ratios were then computed using our compilation of 
atomic parameters, the Kurucz model atmosphere for the Sun \citep{kur93a} 
and this list of Solar equivalent widths. 
We adopt \mtv = 1.0 \kms\ for the Sun. 

The results are listed in Table~\ref{tab_solar}.  
The entries in this table are slightly different than the entries in 
Table~4 of \citet{ram02}, calculated for the M71 analysis, since the 
Solar equivalent widths used in the M71
study were taken from the compilation of \citet{moo66}.
Assuming our atomic parameters, Solar model, Solar \ew\ and analysis
are correct, this calculation should reproduce the current 
compilation of photospheric 
solar abundances of \citet{gre98}.
The difference $\Delta$
between our solar abundances and the photospheric solar 
abundances from \citet{gre98} is listed in column 5 of Table~\ref{tab_solar}.
We ascribe small differences to problems in the absolute scale of the
adopted $gf$ values.  
In general, these differences are reasonably small, 
with the exception of [O/Fe], [Ca/Fe] and [Zr/Fe].  For these three
species, we fail to reproduce the Solar abundance
by more than 0.1 dex.  
The difference we found for [Ca/Fe] ($\Delta=-0.16$ dex)
is almost the same as the correction
factor that needs to be applied to the inverted 
solar $gf$ values from \citet{the89,the90}, none of which were
actually used here since all the detected Ca~I lines for our
M5 sample are included in the NIST database.
The NIST $gf$ scale is within 0.08 dex
of that of \citet{smi81}, but \citet{smi81a} claims
to reproduce the Solar Ca abundance with their $gf$ values.
He achieves this because his Solar \ew\ are considerably
larger than those of \citet{moo66} or those
measured here.  If there is
an error in the NIST Ca I absolute $gf$ value scale, then our procedure
of comparing to our own Solar abundance will remove it exactly
when [Ca/Fe] is calculated for the program stars.
However, if the problem lies in how best to measure the \ew\
of very strong Ca~I Solar lines, our procedure can only remove
the error approximately.  Furthermore
the Solar Ca lines are quite strong, and hence
the inferred Ca abundance is somewhat
dependent on the choice of damping constant, but that is not the
case in general in M5.
The O abundance listed in Table~\ref{tab_solar} corresponds
to the solar value derived from the forbidden lines, which are 
very weak in the Sun.  The three Zr~I lines, which give
$\Delta=+0.29$ dex, 
are also very weak,
but our $gf$ values and Solar \ew\ agree reasonably well with
those of \citet{biemont81}, who carefully analyzed the Solar Zr abundance.
It may be that this is a problem in the details of the temperature
gradient in the outermost layers of our adopted model atmosphere
(see comments in Bi\'emont et al. 1981).  If such an error
is independent of \teff, this is basically the same as having
an error in the scale of the transition probabilities, which would lead
to an incorrect Solar $\epsilon$(Zr), but would not
affect any of our results which are expressed at [Zr/Fe].

La~II and Eu~II are special cases.  The Solar \ew\ for 
the single line of Eu~II and for
those few lines of
La~II that are detected here or in our M71 sample are extremely weak.
There are no $gf$ values in NIST for these lines.  
We were forced to use the transition probabilities from
\citet{corliss62}, adjusted as recommended by 
\citet{arneson77}, for the three La~II lines.  Since the
submission of our M71 paper, a new analysis of the spectrum
of La~II has appeared.  \citet{lawler01a} provide
transition probabilities, hyperfine structure,                     
and a new determination of the Solar La abundance, but only include
one of the three lines used either here or in our analysis of M71 stars.
We therefore adopt the Solar La abundance from \citet{lawler01a},
and scale our $gf$ values by +0.10 dex so that they agree with
the new one for the single line in common, at 6390\AA.  This is
the strongest of the three lines, and the only one seen in most
of our stars in which La~II is detected.  For similar reasons, we adopt
the $gf$ value of \cite{lawler01b} for the single observed
line of Eu~II as well as their Solar Eu abundance of Eu.

We use the solar abundances listed in Table~\ref{tab_solar}, 
derived from our choice of atomic line parameters, to compute the 
iron abundance and abundance ratios for our sample of M5 stars.

\section{STELLAR PARAMETERS}

We follow the philosophy developed for M71 and described in
\citet{coh01}.  We adopt current values
from the on-line database of \citet{har96}
for the apparent distance modulus of M5 at $V$ of 14.31 mag
with a reddening of E(B--V) = 0.03 mag. 
The relative extinction in various passbands is taken from
\citet{coh81} (see also Schlegel, Finkbeiner \& Davis 1998).
Based on the high dispersion analysis of \citet{iva01} for a large
sample of luminous red giants in M5,
we adopt as an initial guess a metallicity for the 
cluster of [Fe/H] = $-$1.0 dex.

We utilize here the grid of predicted broad band colors and
bolometric corrections of \citet{hou00}
based on the MARCS stellar atmosphere code of \citet{gus75}.  
In \cite{coh01} we demonstrated that the Kurucz and MARCS 
predicted colors are essentially
identical, at least for the specific colors used here.

The observed broad band $V$ and $I$ colors for
each program star from the photometric database of \cite{ste98}
and \cite{ste00}, corrected for extinction, are used to determine
\teff.  The set of models with metallicity of $-$1.0 dex, nearest to our
initial estimate of [Fe/H], is used. Table~\ref{tab_ste}
lists the \teff\ thus deduced. 
The reddening to M5 is small, making possible 
extinction variations across the cluster irrelevant.
We assume an random photometric error of 0.02 mag applies to
$V-I$ from \cite{ste00}.  Following \cite{coh01},
this translates into a total uncertainty in \teff\ of 75 K for giants rising
to 150 K for main sequence stars using $V-I$. 
One can obtain a reasonable guess as to the magnitude of possible
systematic errors in Peter Stetson's photometric database for M5
by comparing the first and second versions of his catalog; there
appear to be mean differences in the 
photometric zero points of $\sim0.02$ mag,
corresponding to a mean difference in \teff\ of $\sim$75 K.

We have slightly smoothed the \teff\ for the fainter stars in our
sample by small amounts to ensure that stars at the 
approximately same evolutionary stage have
the approximately the same stellar parameters.

Once an initial guess at \teff\ has been established from a broad
band color, it is possible with minimal assumptions
to evaluate \grav\ using  observational data.
The adopted distance modulus, initial
guess at \teff, and an assumed stellar mass (we adopt 0.8 $M$\subsun\
for the stars in the M5 sample)  are combined with 
the known interstellar absorption, the predictions of the 
model atmosphere grid
for bolometric corrections as well as a broad band observed $V$ mag to
calculate \grav. 
An iterative scheme is used to correct for the small
dependence of the predictions of the model atmosphere grid on
\grav\ itself.  Rapid convergence is achieved.

It is important to note that because of the constraint of
a known distance to M5, the
uncertainty in \grav\ is small, $\le0.1$ dex when comparing
two members of M5.  Propagating an uncertainty of 15\% in the cluster
distance, 5\% in the stellar mass, and a generous 3\% in \teff, 
and ignoring any covariance, leads to
a potential systematic error of $\pm$0.2 dex for \grav.

\subsection{The Spectroscopic Excitation  Temperature}

\begin{figure}
\epsscale{0.7}
% Comment out the following line to embed the PS figure into the manuscript
\plotone{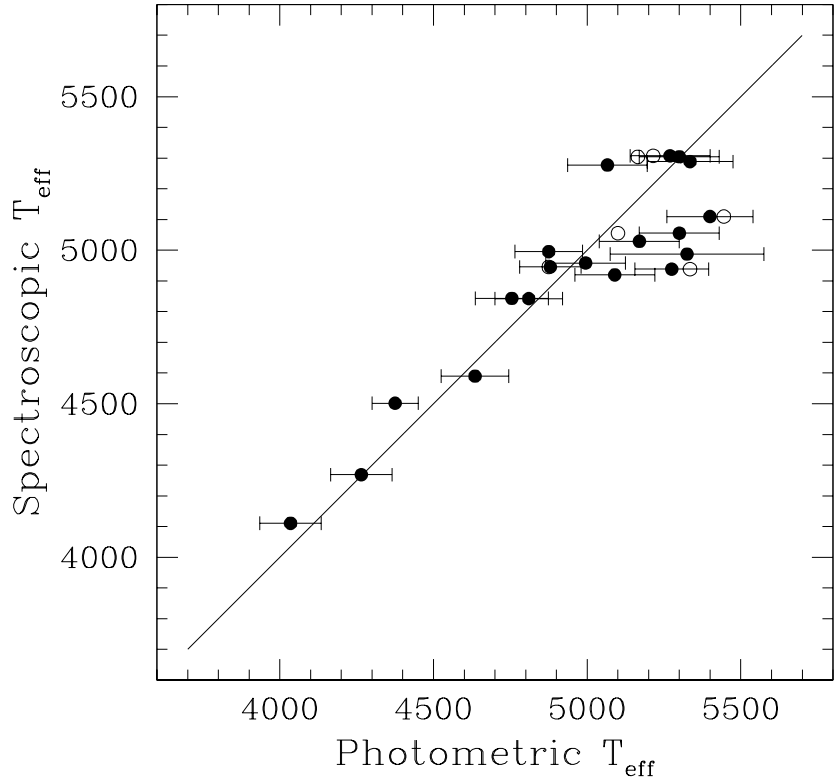}
\caption[teff.ps]{Photometric \teff\ versus spectroscopic (Fe~I excitation) 
\teff\ for the
M5 sample. 
The filled circles show the adopted \teff\, and the open circles
show the original \teff, before smoothing the photometric result
(see text).
The solid line indicates the ideal case when the photometric and
spectroscopic \teff\ are equal. The scatter around the solid line is
about $\pm$130 K. 
\label{teff}}
\end{figure}

The excitation temperature (a spectroscopic measure of \teff) 
of a star can be determined from the observed spectrum
by requiring the derived abundance of an ion with
many observed lines covering a wide range of lower excitation potential
$\chi$ to be independent of $\chi$.
This technique can be applied to 19 of our stars where we have 
detected Fe I lines with sufficient range in $\chi$.
For the determination of the excitation temperature 
we use ``the weak line set'' 
of Fe I lines to ensure that the resulting Fe abundance and derived 
spectroscopic \teff\ 
will be only weakly dependent on the choice of microturbulent velocity.
We find that the spectroscopic \teff\ is in good agreement
with the derived photometric \teff, as shown in Figure~\ref{teff}.
The filled circles show the adopted \teff\, and the open circles
show the original \teff, before smoothing the photometric result,
as described before.
The solid line in Figure~\ref{teff} indicates the ideal case when the 
spectroscopic and the photometric \teff\ are equal.
The scatter around the solid line is about 130 K, which is comparable to the
error of the photometric \teff\ given above.

\subsection{The Microturbulent Velocity}

\begin{figure}
\epsscale{0.7}
% Comment out the following line to embed the PS figure into the manuscript
\plotone{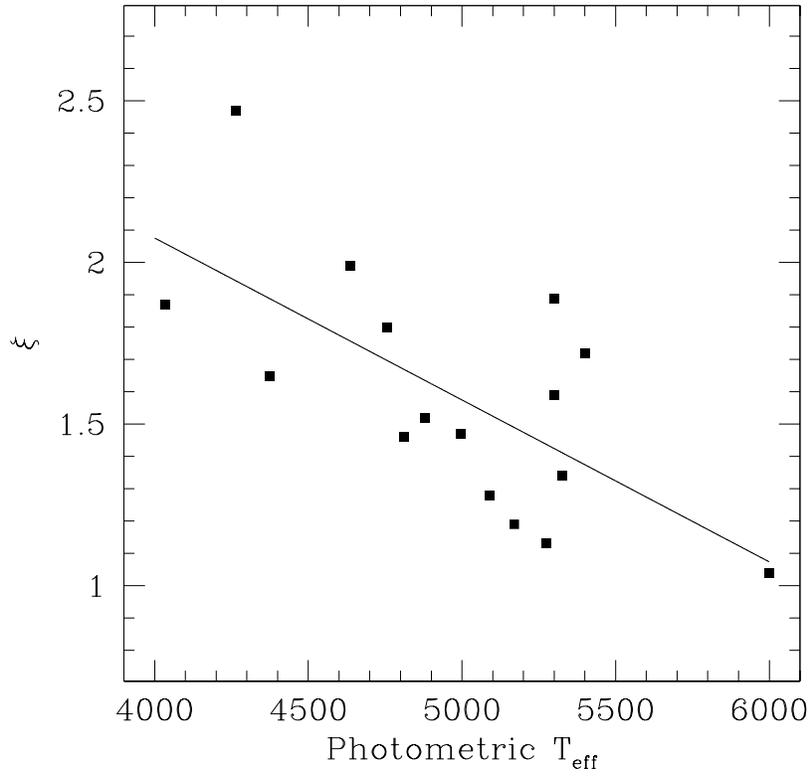}
\caption[microt.ps]{\mtv\ determined for the set of weak Fe I lines
is shown as a function of \teff.
The solid line is the linear fit weighted by the errors.
The scatter around the solid line is about $\pm$0.3 \kms.
\label{microt}}
\end{figure}

The microturbulent velocity (\mtv) of a star can be determined 
spectroscopically by requiring the abundance to be independent of the 
strength of the lines.
We apply this technique for the ``the weak line set'' of Fe I lines.
Only 16 of our stars have enough weak Fe I lines to derive \mtv\
spectroscopically.
The relationship between the determined \mtv\ and the photometric 
\teff\ is shown in Figure~\ref{microt}.
The solid line corresponds to a linear least squares fit to the data,
given by:
$$\xi = 4.08 - 5.01 \times 10^{-4} \times T_{eff} $$
The scatter around the solid line is about 0.3 \kms, which is a reasonable
estimation of the error in \mtv. For the rest of the analysis, we 
adopt for each star the \mtv\ computed from the
\mtv-\teff\ fit.  
The microturbulent velocity used for each star in our 
sample in M5 is listed in 
Table~\ref{tab_ste}.  Unless otherwise specified, we use
in our analysis
the set of all the good atomic lines for each ion
as defined in \S\ref{equiv_widths}.

\section{RESULTS}

Given the derived stellar parameters from Table~\ref{tab_ste}, we 
determined the abundances using the equivalent widths obtained as 
described above.
The abundance analysis is carried out using a current version of the LTE
spectral synthesis program MOOG \citep{sne73}.
We employ the grid of stellar atmospheres from \citet{kur93a} with
a metallicity of \fe\ = --1.0 dex to compute
the abundances of O, Na, Mg, Si, Ca, Sc, Ti, V, Cr, Mn, Fe, Co, Ni,
Cu, Zn, Zr, Ba, La, and Eu using the four stellar atmosphere
models with the closest \teff\ and \grav\ to each star's parameters.
The abundances were interpolated using results from the closest stellar model
atmospheres to the appropriate \teff\ and \grav\ for each star.
We adopt a minimum uncertainty of 0.05 dex in abundance ratios
([X/Fe] or \fe)                              
for ions with less than 10 detected lines in a star, with                       
the minimum lowered to 0.03 dex for ions with more than 10 measured lines     
in a given star.                                                         
If only one line of an ion is detected in a particular star,             
an uncertainty of 0.10 dex is adopted. 
Our final results are not sensitive to small changes in the metallicity
of the model atmosphere (see below).

\subsection{The Iron Abundance \label{sec_iron_abun}}

\begin{figure}
\epsscale{0.9}
% Comment out the following line to embed the PS figure into the manuscript
\plotone{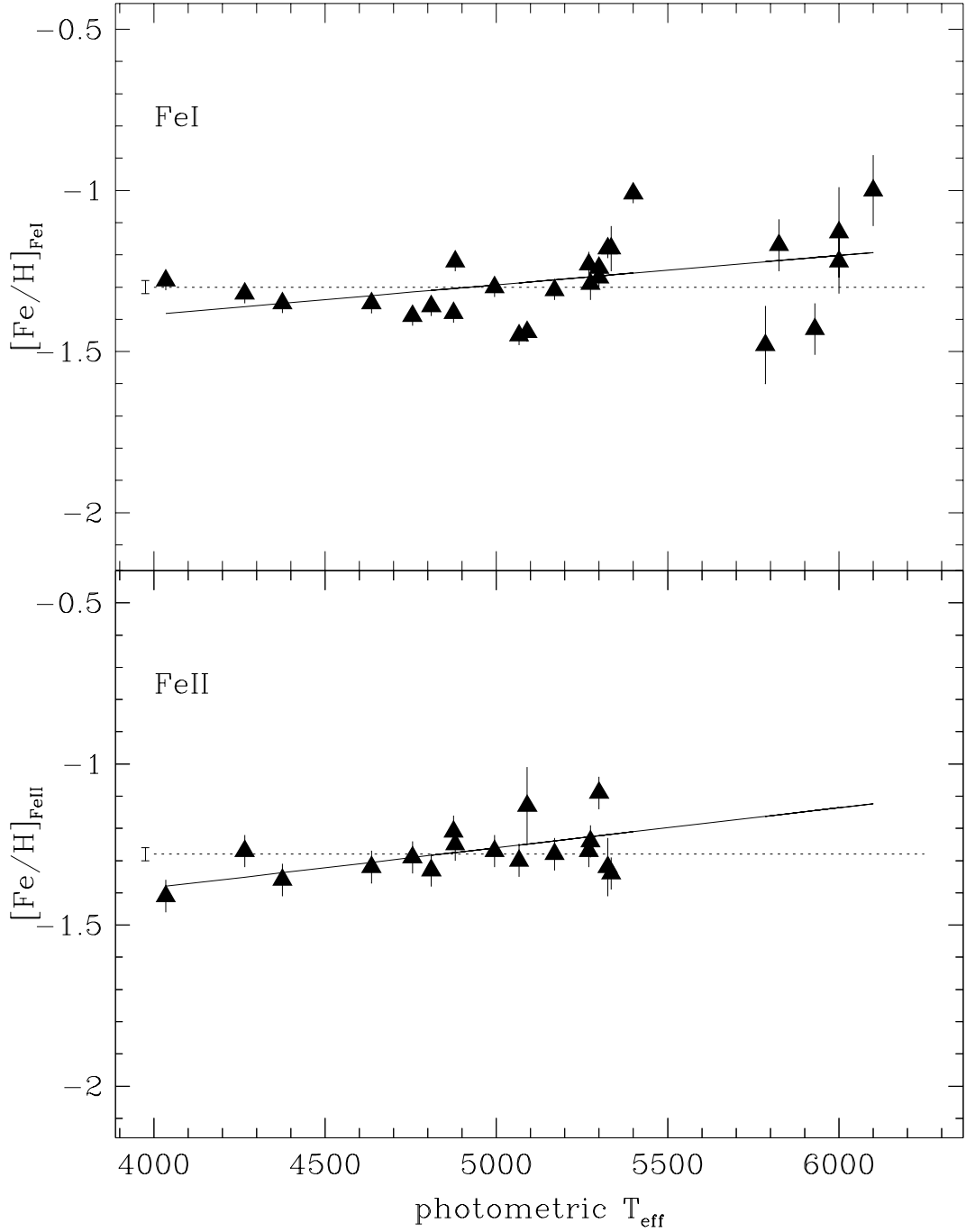}
\caption[fe.ps]{[Fe/H] from Fe I (upper panel) and Fe II (lower panel) against
photometric \teff. The solid lines are linear fits weighted by the errors.
In both cases, \fe\ shows no dependence with \teff. 
The dashed lines indicate the mean \fe\ with their respective error plotted 
as an error bar at 4000 K.
Note that $<$\fe(Fe I)$> = -1.30 \pm 0.02$ and 
$<$\fe(Fe II)$> = -1.28 \pm 0.02$.
\label{fe}}
\end{figure}

The derived abundance \fe\ from Fe I lines 
for each star in our M5 sample is 
listed in column 3 of 
Table~\ref{tab_ratio_a} and plotted against the photometric \teff\ 
in the top panel of Figure~\ref{fe}.
\teff\ is used for the x-axis as a convenient parameter for 
characterizing the position of the stars in the color-magnitude 
diagram as it ranks the stars in luminosity.
The errors listed in column 3 of Table~\ref{tab_ratio_a} correspond 
to the larger 
of the statistical uncertainty, given by the standard deviation of 
the iron abundance from different lines divided by the square root 
of the number of lines of Fe~I used for a particular star, 
or a minimum values based on the number of detected lines specified above.
These errors are lower limits to the actual uncertainties in the 
abundances, since they do not include uncertainties due to the stellar 
parameters nor any systematic effects that might be present.

We evaluate the sensitivity of \fe\ derived from Fe I lines with 
respect to small changes in the equivalent widths and the stellar 
parameters in three cases 4250/1.5/1.8, 5000/3.0/1.5 and 6000/4.0/1.0, 
where the three numbers correspond to \teff/\grav/\mtv.
We estimated the error in the \ew\ to be 10\% for all the lines.
The results are listed in Table~\ref{tab_sensit}, where the range adopted 
for each parameter is representative of its uncertainty.
Our determination of \fe\ from Fe I lines is most sensitive to errors
in \ew\ (a change of $\sim$0.1 dex for a 10\% error in the ~\ew), and has
minimal sensitivity to the choice of metallicity of the model 
atmosphere for plausible changes in \fe\ ($\pm$0.2 dex).
The solid line, shown in the top panel of Figure~\ref{fe}, is a linear fit 
(weighted by the errors) to \fe\ versus \teff.
The slope of the fit is $(+9.1 \pm 5.0) \times 10^{-5}$ dex/K,
which is consistent with \fe\ being constant, independent of \teff\ 
(ie, of luminosity or equivalently position in the color-magnitude diagram),
within a 2$\sigma$ level.
The mean \fe\ weighted by the errors of all 25 stars is $-1.30 \pm 0.02$, 
in very good agreement with earlier determinations \citep{sne92,she96,iva01}.

The determinations of \fe\ from Fe II lines are listed in column 5 of 
Table~\ref{tab_ratio_a} and plotted against the photometric \teff\ 
in the bottom panel of Figure~\ref{fe}.
The errors listed in column 5 of Table~\ref{tab_ratio_a} 
correspond to the larger of the statistical 
uncertainty or to the minimum uncertainty specified
above based on the number of detected lines.
We evaluate the sensitivity of \fe\ derived from Fe II lines with respect to
small changes in the equivalent widths and the stellar parameters in the 
same manner as the sensitivity of \fe\ from Fe I lines. 
The results are listed in Table~\ref{tab_sensit}, where the 
range adopted for each parameter is representative of its uncertainty.
We see a stronger sensitivity on the stellar parameters from the
Fe II lines than from the Fe I lines.
The \fe\ determination from Fe II lines is most sensitive to the systematic
error in \grav\
(note that the internal uncertainty in \grav\ is smaller, $\sim$ 0.1 dex),
as well as to \teff\ among the coolest M5 giants.
The sensitivity on the choice of metallicity of the model atmosphere
is again small for reasonable changes in metallicity.
The solid line, shown in the bottom panel of Figure~\ref{fe}, is a linear 
fit weighted by the errors of \fe\ versus \teff.
The slope of the fit is $(+1.2 \pm 0.5) \times 10^{-4}$ dex/K,
which is essentially identical to the small, not statistically significant
slope obtained from the Fe~I lines. 
The mean \fe\  weighted by the errors for the 17 M5 stars with 
detected Fe II 
lines is $-1.28 \pm 0.02$, in very good agreement with our result from 
Fe I lines and earlier determinations \citep{sne92,she96,iva01}.

The iron abundance could be affected by departures from LTE. The main NLTE
effect in late-type stars arises from overionization of electron donor
metals by ultraviolet radiation \citep{aum75}.
\citet{gra99} and \citet{the99} studied NLTE effects in Fe
abundances in metal-poor late-type stars.
\citet{gra99} found that NLTE corrections for Fe lines are very small in
dwarfs of any \teff, and only small corrections ($<$ 0.1 dex) are expected for
stars on the red giant branch.
\citet{the99} found that NLTE corrections become more important as [Fe/H]
decreases, being about 0.2 dex for stars with \fe $\sim -$1.25 dex, and that
ionized lines are not significantly affected by NLTE.  Very recently,
\cite{geh01a} and \cite{geh01b} have carefully calculated the kinetic
equilibrium of Fe, and present in \cite{kor02}
a critique of earlier calculations.  They suggest non-LTE corrections intermediate between
the above sets of values are appropriate 
for Fe~I.

One way to explore possible NLTE effects present in our data is by comparing
the results from Fe I and Fe II lines. The mean difference between \fe\ 
from Fe II and Fe I lines is 0.02 $\pm$ 0.18.
The slope of the relationship between 
(\fe$_{\rm{FeII}}$-\fe$_{\rm{FeI}}$) vs.~\teff\ 
is $(+0.6 \pm 1.0)\times 10^{-4}$ dex/K, which is nearly flat. 
We conclude that NLTE effects are negligible in our iron abundance 
determination, with a maximum change of 0.12$\pm0.2$ dex in
the Fe ionization equilibrium from the tip of the RGB to the main
sequence turnoff in M5.

\subsection{Abundance Ratios}

The abundance ratios, with the exception of [O/Fe], [Si/Fe] and [Zn/Fe], 
are computed using the iron abundance from Fe I lines                    
and our solar abundance ratios from Table~\ref{tab_solar}.               
Given their high excitation potentials, the abundance ratios for the     
Si I and Zn I lines were computed using the [Fe/H] from Fe II lines.     
In the \teff\ range of our sample of stars in M5, most of the iron is 
in the form of Fe II and most of the oxygen is in the form of O I, so both      
species behave similarly for small changes in the atmospheric parameters.
For this reason, we computed the abundance ratio of O using the          
Fe II lines as well.                                                     
The computed abundance ratios are listed in                              
Tables ~\ref{tab_ratio_a} - ~\ref{tab_ratio_d}.                          
The error listed in Tables ~\ref{tab_ratio_a} - ~\ref{tab_ratio_d}       
for each ion corresponds to larger of the statistical uncertainty in the        
mean abundance for the lines of that ion detected in a particular star,
i.e. to the standard deviation 
within our sample of stars divided by the square root of the number of 
stars for which an abundance was derived for that ion, or to
the minimum uncertainty specified above, depending on the number
of detected lines.  

The O lines were detected in nine of our M5 stars;  we provide 
constraining
upper limits on their equivalent widths in other six stars. Among the
nine stars with clear detections, we were able to measure both the permitted 
and forbidden lines in five of them. We find no significant difference 
between the oxygen abundance derived from permitted and forbidden lines; 
the [O/Fe] listed in Table~\ref{tab_ratio_a} corresponds to the average 
results from all the lines detected in each star.  

In M71, however,
(see \S3 of Ram\'{\i}rez \& Cohen 2002)
we did see a difference between the O abundance deduced from
the permitted and from the forbidden lines.  R. P. Kraft suggested to
us that this might be the consequence of adoption of a reddening 
value for
M71 which is slightly too small. (We used E(B$-$V)=0.25 mag.)
This would result in an underestimation of
\teff\ for each M71 star.  An increase in the reddening 
for M71 of only 0.03 mag would produce a change of $\approx$100K in 
the resulting \teff\ obtained from the broad band colors
$V-I$, $V-J$ and $V-K$.  The high excitation potential
of the permitted O lines leads to a very high sensitivity to \teff\
(see Table~\ref{tab_sensit}).
We have checked that a 100K increase in \teff\ will in fact
eliminate the trend displayed in Figure 1 of \cite{ram02}
and produce approximate agreement between the two independent
determinations of the O abundance for each star in our sample
in M71 for which both permitted and forbidden O lines could be 
measured.  If one chooses to
adopt this higher value for the reddening of M71,  
small changes in the abundance ratios for 
the M71 stars then follow, which are given in Table 6 of \cite{ram02}.  
At present we lack an accurate
determination of the reddening of M71, which might in principle be
obtained directly from high precision spectra of the brighter M71
stars without recourse to photometry.
Fortunately, the reddening to M5 is much smaller.

The Na D lines were detected in all 25 stars in our sample. Given 
the low reddening of M5,
the stellar Na D lines are relatively free of contamination by
interstellar features, and the $v_r$ of M5 
shifts the night sky emission lines
away from the Na D lines of the cluster stars.  Thus the Na D
lines could be and were included in the analysis.
(This was not possible for M71.)
We compare the resulting abundances for
the 20 stars where both the Na D lines and the 5682,5688 \AA\ 
Na doublet were detected.
There is a correlation between the sodium abundance derived from
the D lines and the one derived from the weak Na lines which does
not quite correspond to the desired one of equality.  
To be consistent, the sodium abundance derived from the D lines was
put on the same scale as the sodium abundance derived from the other
Na lines, before taking the average.

The spectra of the six main sequence stars in
our sample in M5 do not have high signal-to-noise
ratios and the lines are in general weak.  In a few cases where no line
of an ion could be detected in any individual spectrum, we summed the spectra,
and this enabled detection of a small number of additional lines.  This was
done for the O I triplet, the Mg I line at 5528\AA, 
and two Sc II lines.  
Their equivalent widths and determined abundance ratios are 
listed in Tables~\ref{tab_ew} and ~\ref{tab_ratio_a} -- ~\ref{tab_ratio_b},
respectively, under the entry "$<$MS$>$".
In each of these cases, the temperature sensitivity of the lines
increases their strength in these hotter low luminosity M5 stars. 

\begin{figure}
\epsscale{0.7}
% Comment out the following line to embed the PS figure into the manuscript
\plotone{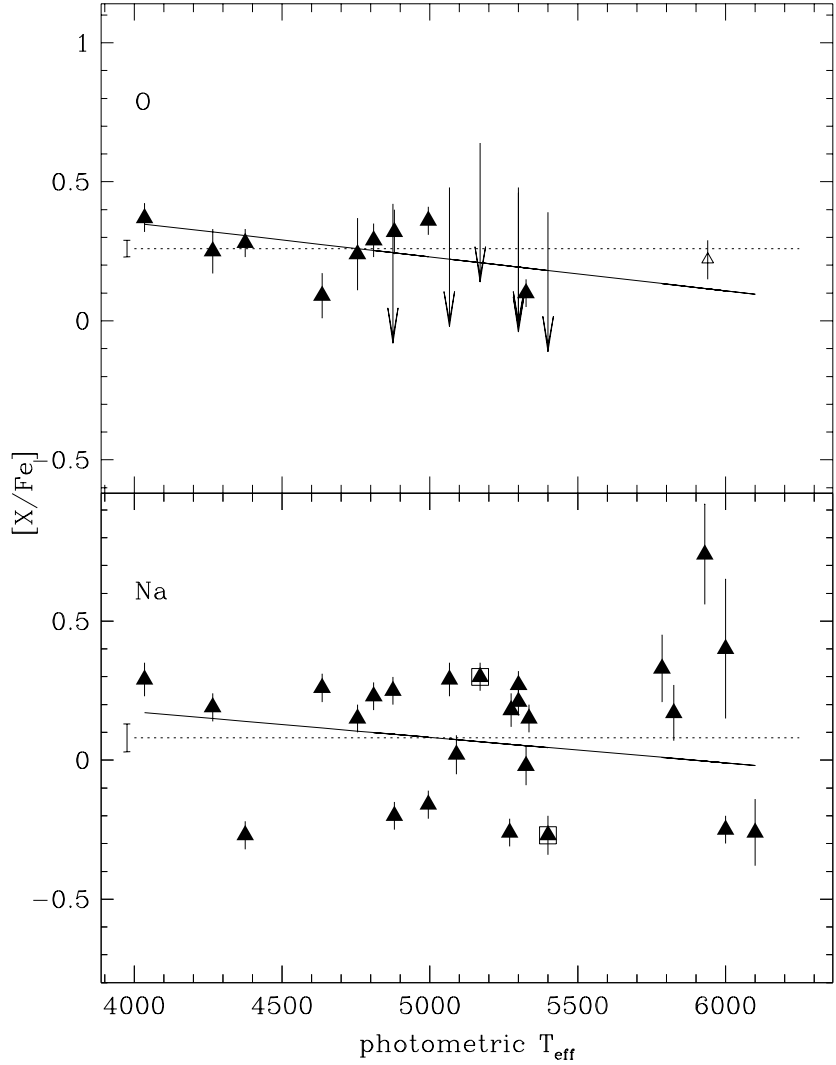}
\figcaption[light.ps]{Abundance ratios of 
O and Na 
with respect to Fe against \teff.  
The solid line is a linear fit weighted by the errors.
The dashed line indicates the mean abundance ratio with its respective 
error plotted as an error bar at 4000 K.
The open triangle corresponds to the abundance determined from the summed
spectra of the six main sequence stars.
Arrows represent upper limits for the oxygen abundance ratio.
Stars G18450\_0453 and G18564\_0457, part of whose spectra are shown in 
Figure~\ref{na_spec}, are marked with open squares in the [Na/Fe] panel.
\label{light}}
\end{figure}

\begin{figure}
\epsscale{0.7}
% Comment out the following line to embed the PS figure into the manuscript
\plotone{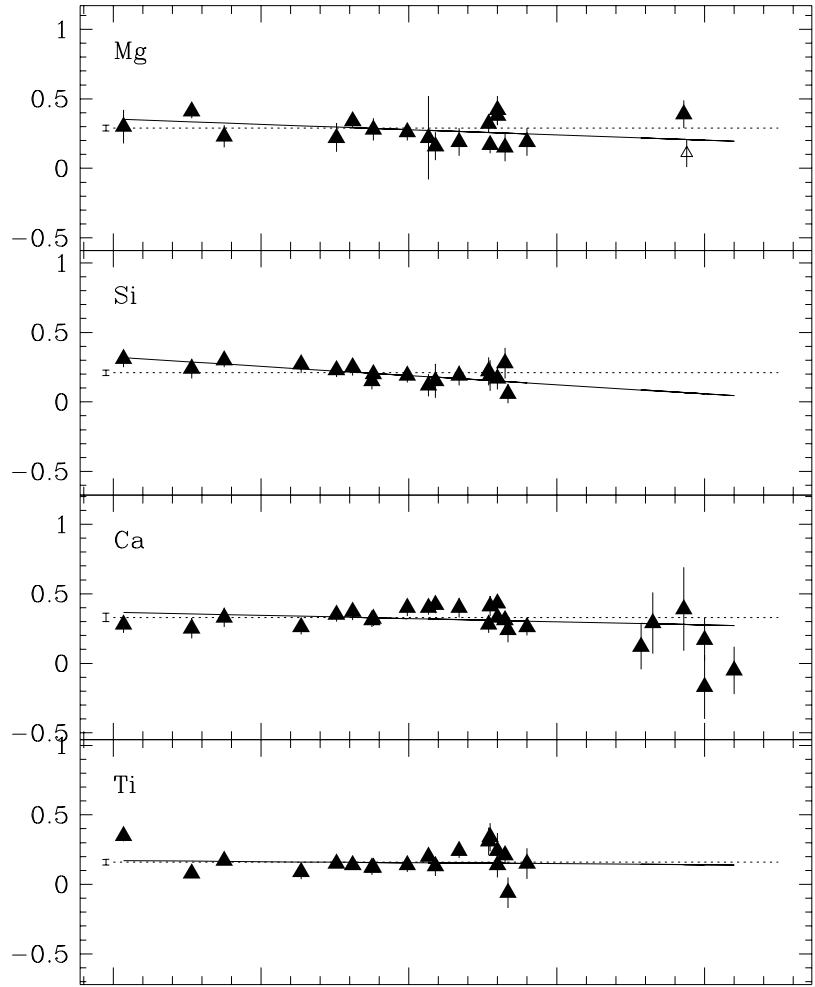}
\figcaption[alpha.ps]{Abundance ratios of the $\alpha-$elements 
Mg, Si, Ca and Ti
with respect to Fe against \teff.
The symbols are the same as in Figure~\ref{light}.  
\label{alpha}}
\end{figure}

\begin{figure}
\epsscale{0.7}
% Comment out the following line to embed the PS figure into the manuscript
\plotone{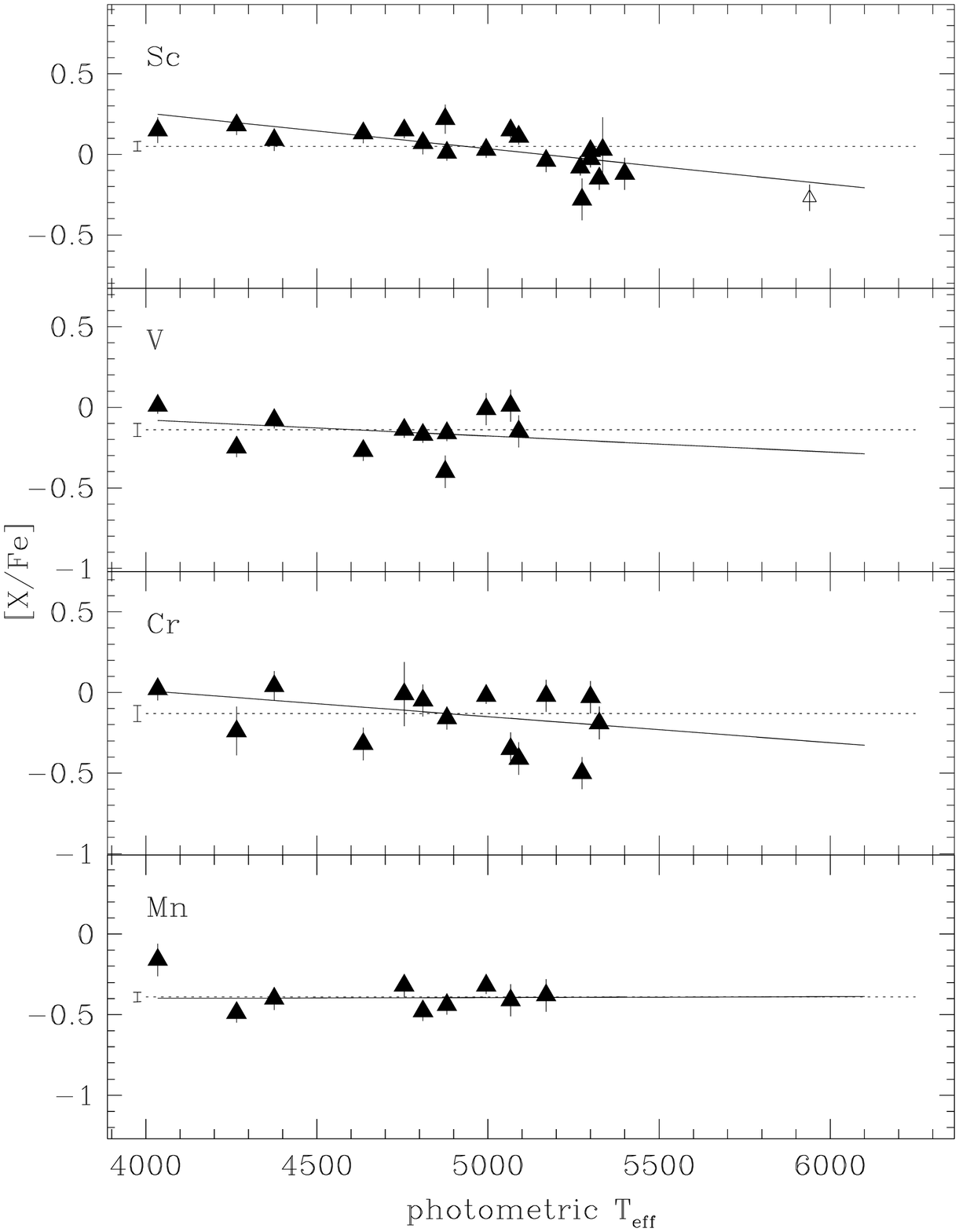}
\figcaption[ironp1.ps]{Abundance ratios of the iron peak elements 
Sc, V, Cr, and Mn
with respect to Fe against \teff.
The symbols are the same as in Figure~\ref{light}.
\label{ironp1}}
\end{figure}

\begin{figure}
\epsscale{0.7}
% Comment out the following line to embed the PS figure into the manuscript
\plotone{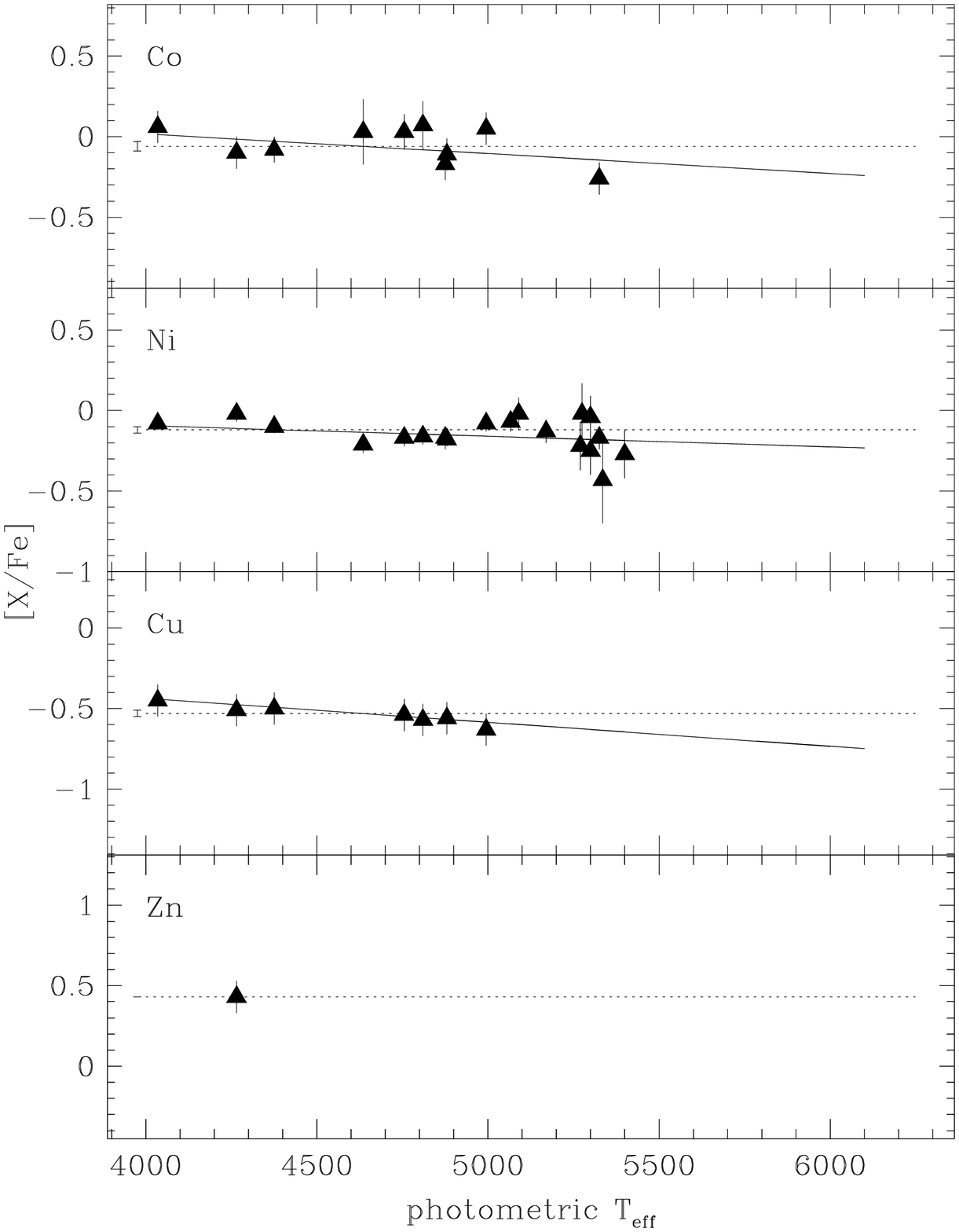}
\figcaption[ironp2.ps]{Abundance ratios of the iron peak elements 
Co, Ni, Cu, and Zn
with respect to Fe against \teff.
The symbols are the same as in Figure~\ref{light}.
\label{ironp2}}
\end{figure}

\begin{figure}
\epsscale{0.7}
% Comment out the following line to embed the PS figure into the manuscript
\plotone{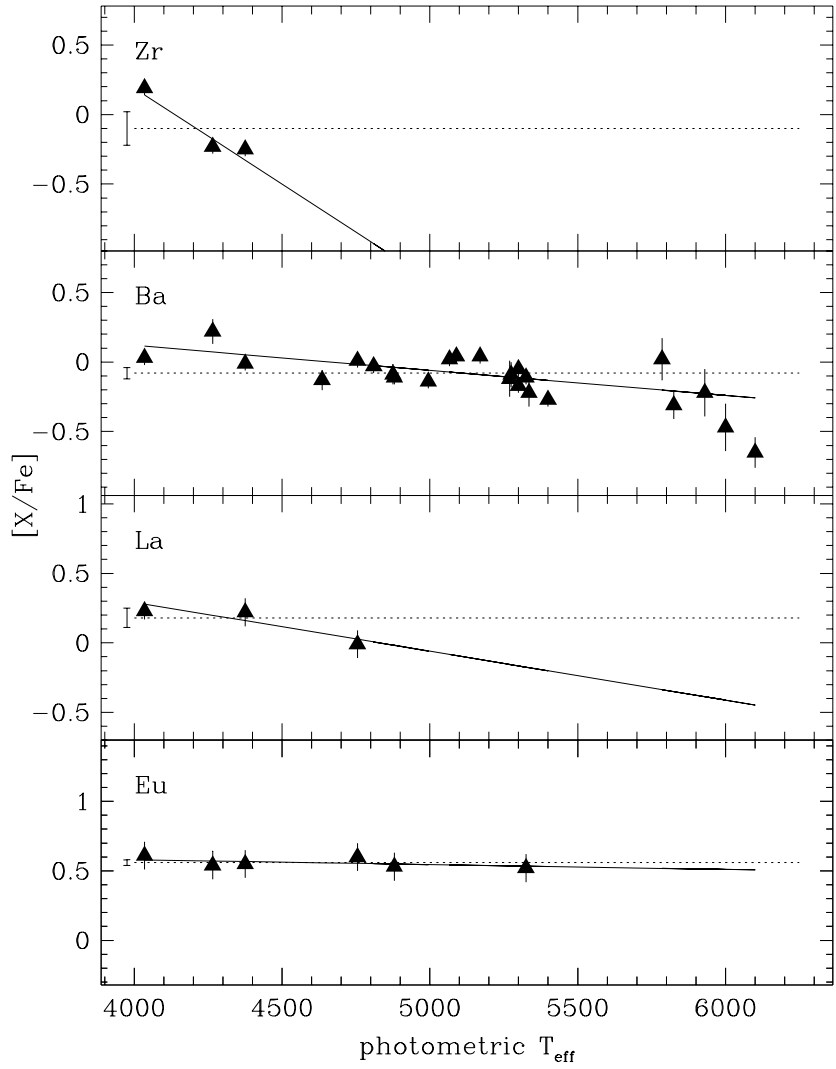}
\figcaption[neutron.ps]{Abundance ratios of the neutron capture elements 
Zr, Ba, La, and Eu
with respect to Fe against \teff.
The symbols are the same as in Figure~\ref{light}.
\label{neutron}}
\end{figure}

The abundance ratios for each star in our M5 sample are plotted against 
the photometric \teff\ in Figures~\ref{light} to \ref{neutron}. 
The error bars shown in Figures~\ref{light} to~\ref{neutron} 
are those described above.
The solid line, shown in Figures~\ref{light} to~\ref{neutron}, is a 
linear fit weighted by the errors of the respective abundance ratio 
versus \teff.
The dashed line shown in these figures indicates the mean 
abundance ratio and its respective error plotted as an error bar at 4000 K.
In the top panel of Figure~\ref{light} the arrows correspond to the upper
limit in [O/Fe] for six stars in our sample.
An open triangle indicates the mean abundance measured in the summed
spectrum of the six main sequence stars.

We estimate the sensitivity of abundances with respect to
small changes in the equivalent widths and the stellar parameters 
in the same manner as the sensitivity of \fe\ from Fe I lines. 
The results are listed in Table~\ref{tab_sensit}, where the
range adopted for each parameter is representative of its uncertainty.
The \ew\ of the lines of
Ti~I, V~I and Zr~I have the strongest dependence on \teff\
as most of the detected lines from these neutral ions
have low excitation potentials.  Due to its strong lines,
Ba~II has the strongest dependence on \mtv.

The mean abundance ratios and their errors are listed in 
Table~\ref{tab_mean}. 
The error of the mean for an element $X$ corresponds to the standard 
deviation of the abundance ratio [$X$/Fe] of the sample of stars 
in which lines of $X$ were detected, 
divided by the square root of the number of such stars. 
The error of the mean computed this way represent the formal 
statistical uncertainty in the mean abundance considering only the internal
errors; it does not include any possible systematic
errors that may arise during the analysis.
The standard deviation about the mean abundance ratio, 
$\sigma_{obs}$, is a measure of the scatter 
of the abundance ratio for a particular ion within the sample of M5 stars.  
In order to quantify the abundance ratio variations within our sample of 
M5 stars we have to compare the measure of the scatter with the predicted
error from the stellar parameters and the measurement of the \ew.
We estimated the predicted error, $\sigma_{pred}$, in the same manner as
in \citet{ram02}.
Our $\sigma_{pred}$ ignores covariance among the error terms, which
is discussed in detail by \citet{joh02}.  She shows that these
additional terms are fairly small, and will be even smaller in our case, 
as we have determined \grav\ using cluster distances rather than through
the ionization equilibria of Fe.
The general small trends seen in Figures~\ref{light} to \ref{neutron} 
of [$X$/Fe] slightly increasing toward cooler \teff,
a trend also seen in our analysis of M71 \citep{ram01,ram02},
may result from ignoring the covariance terms (see Johnson 2002).
The predicted errors for each ion are listed in column 5 of 
Table~\ref{tab_mean}.

\begin{figure}
\epsscale{0.7}
% Comment out the following line to embed the PS figure into the manuscript
\plotone{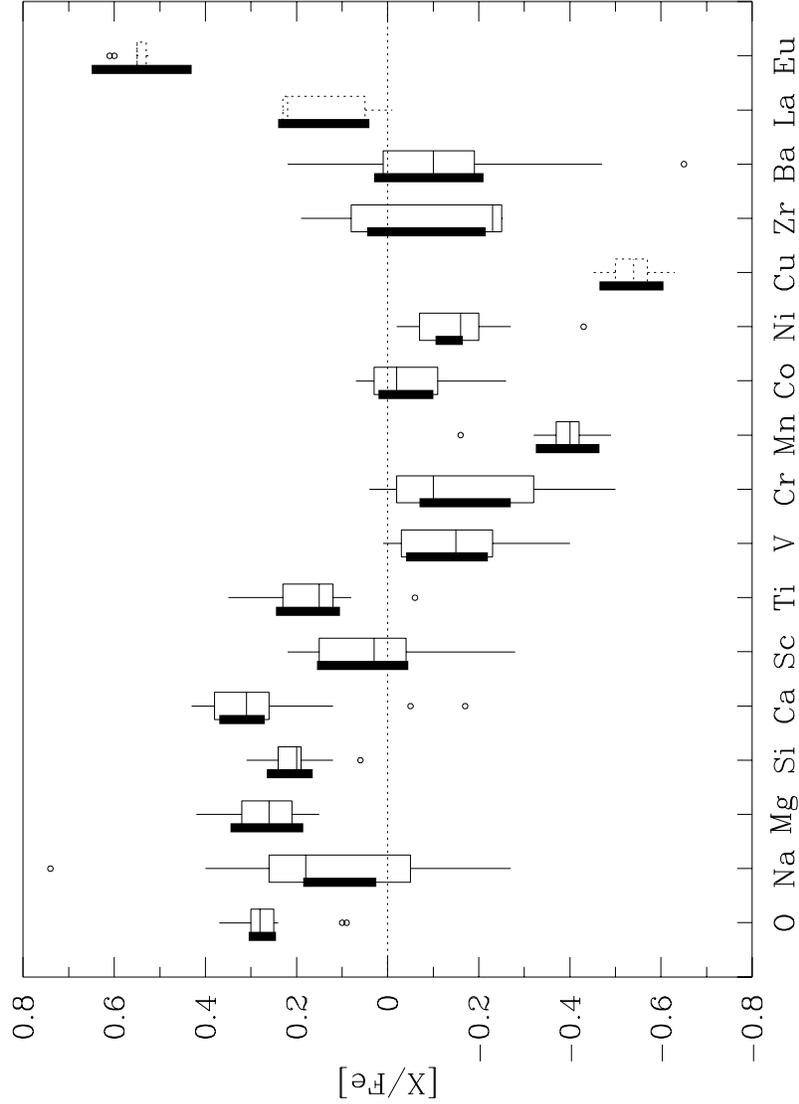}
\figcaption[summ_fig.ps]
{Summary of abundance ratios in M5. Each abundance ratio is plotted with a box 
whose central horizontal line is the median abundance ratio, the bottom 
and the top shows its inter--quartile range, the vertical lines coming 
out of the box mark the position of the adjacent points of the sample, 
and the outliers are plotted as open circles. Boxes constructed
with dashed lines denote elements where only one line per star was
observed. The thick line on the left 
side of the box is the predicted error (expected for the 
inter--quartile range) which includes the dependence on the stellar
parameters and the equivalent width determination.
\label{summ_fig}}
\end{figure}

A summary of the abundance ratios for our M5 sample
is shown in Figure~\ref{summ_fig}. 
The results for each element are depicted as a box 
plot defined by \cite{tuk77} (see also Cleveland 1993).
The central horizontal line in each box is the median abundance 
ratio for all the M5 stars 
included, while the bottom and the top shows its inter--quartile 
range, the vertical lines coming out of the box mark the position 
of the adjacent points of the sample, and the outliers are plotted 
as open circles. 
The  boxes drawn with dotted lines correspond to elements with 
abundances computed from only one line in each star and hence 
are more uncertain.
The thick line on the left side of the box is the predicted 
1$\sigma$ rms error scaled to correspond to the $\pm$25\% 
inter--quartile range. 
Figure~\ref{summ_fig} suggests that the element where we are most likely 
to see star-to-star variations in our M5 sample
is Na;  these variations were in fact detected and
will be discussed in detail in \S\ref{section_o}. 

\section{DISCUSSION}

\subsection{[Fe/H] and Diffusion of Heavy Elements \label{section_diffusion}}

Our \fe\ abundance results provide further evidence that the iron 
abundance, derived from both Fe I and Fe II lines, is independent of 
\teff, and equivalently luminosity, within the globular cluster M5,
from the top of the giant branch to near the main sequence.
Our result in M5 is in agreement with our previous analysis of a 
sample of M71 stars covering a 
similar range in luminosity \citep{ram01}.
The work of \citet{gra01} also supports our \fe\ abundance result.
They presented abundances from high dispersion spectra from the VLT of
stars in NGC 6397 and NGC 6752, finding that \fe\ obtained for stars at 
the base of the subgiant branch agrees within a few percent with the 
\fe\ obtained for stars at the main sequence turnoff.
Our results, in both M71 and M5, and those of \citet{gra01}, appear to 
be in disagreement with inhomogeneities in \fe\ found earlier by 
\citet{kin98} in a small sample of stars in M92.
These inhomogeneities may rise from the comparison of data sets analyzed 
by different groups, who may have determined the stellar parameters and 
performed the abundance determinations in a different way.

It is clear from helioseismology that atomic diffusion (primarily 
of He) occurs in the Sun (Basu, Pinssoneault \& Bahcall 2000 
and references therein).  Theoretical investigations
of heavy element diffusion in metal poor stars by
\cite{mic81}, \cite{cas97}, \cite{sal00}, \cite{sal01} and
\cite{cha01}, among others, 
suggest that it should be important in metal poor stars
at the main sequence turnoff, while at the level of the RGB, the
presence of deep
surface convection zones should inhibit such processes. Such
processes should manifest themselves in most clearly in globular cluster
stars due
to their old age, giving lots of time for such phenomena
to develop, and presumed identical initial chemical composition. They would
produce a trend of decreasing abundance with \teff\ in our
data, of an amplitude of $\sim$0.2 dex for Fe between the
RGB and the main sequence turnoff.  The very
recent set of models including radiative acceleration as well as
gravitational settling by
\cite{ric02} suggest that  Fe will rise in metal poor main sequence
stars rather than sink, reaching an excess of a factor of 5
at [Fe/H]$ = -2.3$ dex just before a star evolves off the main
sequence to become a subgiant.  Similar large enhancements
are predicted for the elements Al to Ca, with the enhancements
depending on the age of the star.  
These theoretical predictions are highly
dependent on the details of the modeling of the relevant physical processes.

Although gravitational settling and radiative levitation are clearly
detected in hot horizontal branch stars in globular clusters
\citep{beh99,beh00}, we continue
to see no evidence for such effects in our sample of 
red giant branch, subgiants, and main sequence globular cluster stars, 
either in Fe I or in
any of the abundance ratios discussed below.  To within the precision
of our measurements, there is no evidence
for any change in abundances with decreasing \teff, i.e. with
decreasing luminosity, between the tip of the RGB and the main sequence
turnoff. In agreement with other recent studies, particularly
that of \cite{gra01} and \cite{bon02} in NGC~6397 and 
our previous work \citep{ram02}
in M71, we suggest that
some process must provide an additional turbulence in low metallicity
main sequence stars which inhibits the predicted diffusion of heavy
elements.

\subsection{Fe-peak elements}

The abundance ratios of [Sc/Fe], [V/Fe], [Cr/Fe], [Mn/Fe], [Co/Fe], 
and [Ni/Fe] follow the behavior of iron as expected, showing no 
significant trend with \teff, and less or similar scatter around the 
mean than the predicted error.  The exception is Cr~I, whose 
scatter appears 
unexpectedly large in Fig.~\ref{ironp1} and in 
Table~\ref{tab_mean}.  However, 
although nine lines of this ion were detected
in the coolest and most luminous star in our sample,
only a single Cr~I line
could be detected in half of the 14 M5 stars in which we
could detect any Cr~I lines.  
The mean abundance ratios of Sc, V, and Ni 
are consistent with the earlier results of \citet{sne92}, who analyzed 
high resolution spectra of 13 giant stars in M5. Our iron--peak 
abundance ratios are also in agreement with those
results obtained by \citet{iva01}, who analyzed high resolution 
spectra of 36 giants in M5.  A comparison of our abundance ratios
with those of these two previous studies for all elements analyzed
by them
is given in Table~\ref{tab_comp}, which will be discussed in detail
in \S\ref{ivans_comp}.

\subsection{Neutron capture elements}

The abundance ratios of the neutron capture elements, Zr, Ba, La, 
and Eu, show no significant trend with \teff, and scatter 
around the mean similar in size to the predicted error.
The mean abundance ratios of Ba, La (see \S\ref{ivans_comp} 
for details), and Eu
are consistent with the earlier results of \citet{iva01} listed
in Table~\ref{tab_comp}.
The abundances of Ba, La, and Eu are overabundant relative to Fe, as
is seen in other globular clusters such as M4 \citep{iva99} and
M15 \citep{sne97}. 

The mean [Ba/Eu] ratio of --0.64 is consistent with values
observed in halo stars of similar [Fe/H] \citep{bur00,gra94}.
At [Fe/H] $\sim$ --1.3 dex, as reviewed by \cite{bur00},
Ba is a neutron capture element synthesized 
through $s-$process reactions that occur mainly in low mass asymptotic 
giant branch stars, while Eu is exclusively an $r-$process element.

There are seven stable isotopes of Ba. 
The Ba isotope ratios in a star depend on the 
relative contribution of $s$-process to $r$-process production\footnote{We 
ignore two Ba isotopes of very low abundance,
($^{130}$Ba and $^{132}$Ba), which are formed in neither the
$r$ nor the $s$-process.}.
A pure $r$-process isotopic distribution is assumed by \cite{macwilliam98},
who calculated the hfs splittings we use.  
As first described by \cite{rut78},
$^{135}$Ba and $^{137}$Ba have components that are shifted both
to longer and to shorter wavelengths from the line center,
defined roughly by the position of the components from the
even isotopes $^{134}$Ba, $^{136}$Ba and $^{138}$Ba.
In principle, as suggested
by \cite{mas99}, since the hfs splittings are different for
different odd and even isotopes, the deduced abundances 
from a set of Ba~II lines
can be used to infer the isotopic distribution and hence
constrain the Ba production mechanism(s).  In practice, however,
as described by \cite{lambert02},
this is very difficult, even with very high precision spectra with
very high signal-to-noise ratio.
The issue of the isotopic distribution is not relevant for Eu,
where $r$-process production dominates under all circumstances.

The Zr~I lines, which were detected automatically in the
three coolest M5 giants, have been checked
by hand.  The extremely strong Zr~I lines in the luminous giant M5 IV-81
are real.  In this temperature range, the strength of the
Zr~I lines increases rapidly as \teff\ decreases (see
Table~\ref{tab_sensit}).
An overestimate of \teff\ by 200 K error for this star
alone would be required
to make the Zr abundance of this star agree with that of the other
stars in M5 in which this element was detected.  Star IV-81 is
included in the sample of \cite{iva01}, where it is assigned
\teff=3950 K, 85 K cooler than the effective temperature adopted here.
The evidence is inconclusive, but it is interesting to note that
we had similar concerns regarding the possible existence 
of star-to-star variations of Zr in our M71 sample (see 
Ram\'{\i}rez \etal\ 2001).

\subsection{$\alpha-$elements}

We find that the $\alpha-$elements Mg, Ca, Si and Ti are overabundant 
relative to Fe. 
Our mean abundance ratios for Ti and Si are
similar to the results of \citet{sne92} for 13 M5 giant stars, and
also similar to the abundance ratios derived by \citet{iva01}
for 36 giant stars, as listed in Table~\ref{tab_comp}.
Our $<$[Ca/Fe]$>$ is higher than the values of \citet{sne92} 
and \citet{iva01}
The $\alpha-$element abundance ratios show no significant trend with \teff, 
and low scatter around the mean.

[Mg/Fe] is know to vary among bright giant stars in some metal poor 
globular clusters, but no such variation is detected in our sample of
stars in M5.
In NGC 6752 \citep{gra01}, M13 \citep{kra93,she96} and M15 \citep{sne97}, 
[Mg/Fe] shows a star-to-star range in abundance of more than 1 dex.
The mean of the abundance ratio [Mg/Fe] we find in our sample of
25 stars in M5 as well as the scatter about this mean are 
similar to the results of \citet{iva01}.
Furthermore, our comparison between the observed scatter and the 
predicted error of [Mg/Fe] given in Table~\ref{tab_mean} indicates no 
sign of star-to-star variation of magnesium in M5.
[Mg/Fe] showed similar behavior in our previous study 
sample of M71 stars \citep{ram02}; in particular, no variation in Mg
abundance was detected in that cluster either.

\subsection{Sodium and Oxygen \label{section_o}}

\begin{figure}
\epsscale{0.7}
% Comment out the following line to embed the PS figure into the manuscript
\plotone{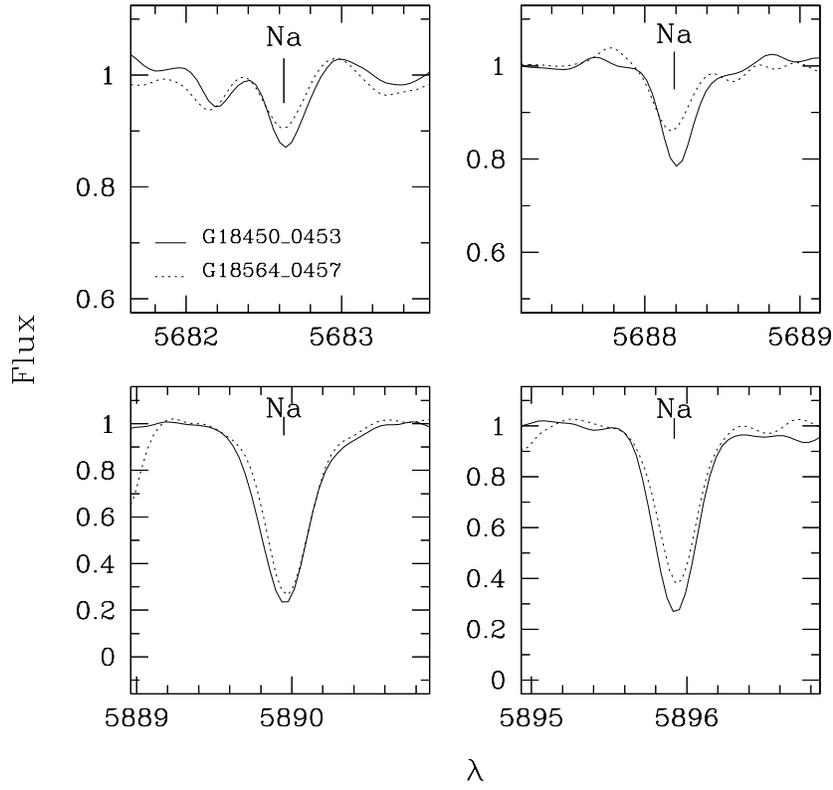}
\figcaption[na_spec.ps]{Comparison of the strength of four 
Na I between two stars of similar effective temperatures, 
G18450\_0453 (5170 K, [Na/Fe]=+0.30) and  
G18564\_0457 (5400 K, [Na/Fe]=--0.27). 
The scatter shown by [Na/Fe] is due to real abundance variations 
among stars of similar \teff.
\label{na_spec}}
\end{figure}

The sodium abundance ratios in our sample of stars in M5 behave 
differently than the abundance ratios of all other elements 
included in this paper. 
The observed scatter for [Na/Fe] is more than
twice as large as the value of $\sigma_{pred}$ given in  
Table~\ref{tab_mean}.
This strongly suggests that the
scatter in [Na/Fe] shown in Figure~\ref{light}
arises from star-to-star abundance variations of Na
within the 25 stars of our M5 sample.
In Figure~\ref{na_spec} we compare the strength of the Na I 
lines between two stars of similar stellar parameters.
The star G18450\_0453, whose slightly smoothed spectrum is shown 
as a solid curve, has a high [Na/Fe]
(+0.30 dex), and the star G18564\_0457, whose slightly smoothed 
spectrum is shown as a dotted curve,
has a low [Na/Fe] (--0.27 dex).  
These two stars are marked with open squares in the bottom
panel of Figure~\ref{light}.
This figure demonstrates again that the higher scatter seen in [Na/Fe]
is due to star-to-star abundance variations.
Note that both of these stars are red giants more than a 
magnitude fainter than the horizontal branch.

We have succeeded in reliably detecting O lines in only nine stars of 
our M5 sample, a reflection of the low metallicity of M5 compared to M71.
Among those nine stars, the oxygen abundance ratio is nearly constant 
and its statistical scatter is similar to the predicted error. 
The [O/Fe] among those nine stars ranges from 0.37 dex to 0.09 dex, 
which is significantly smaller 
than the range from 0.47 dex to --0.55 dex obtained 
in the analysis of 36 giants from \citet{iva01}. 
The O-poor and Na-rich stars found by \citet{iva01} are not
present in our sample of nine M5 stars, but may be hidden among
the six stars with upper limits in [O/Fe] or in the rest of
the sample where no useful upper limit could be obtained.

We have summed the spectra of the six main sequence stars 
to detect more lines at this low luminosity. 
We were thus able to measure each of the three lines of the O I triplet
and derived an abundance from them. The value that we obtain
is [O/Fe]$_{\rm MS}$=+0.22$\pm$0.07 (Table~\ref{tab_ratio_a}), 
which is very consistent with the mean oxygen abundance ratio of
the rest of the M5 sample.

\begin{figure}
\epsscale{0.7}
% Comment out the following line to embed the PS figure into the manuscript
\plotone{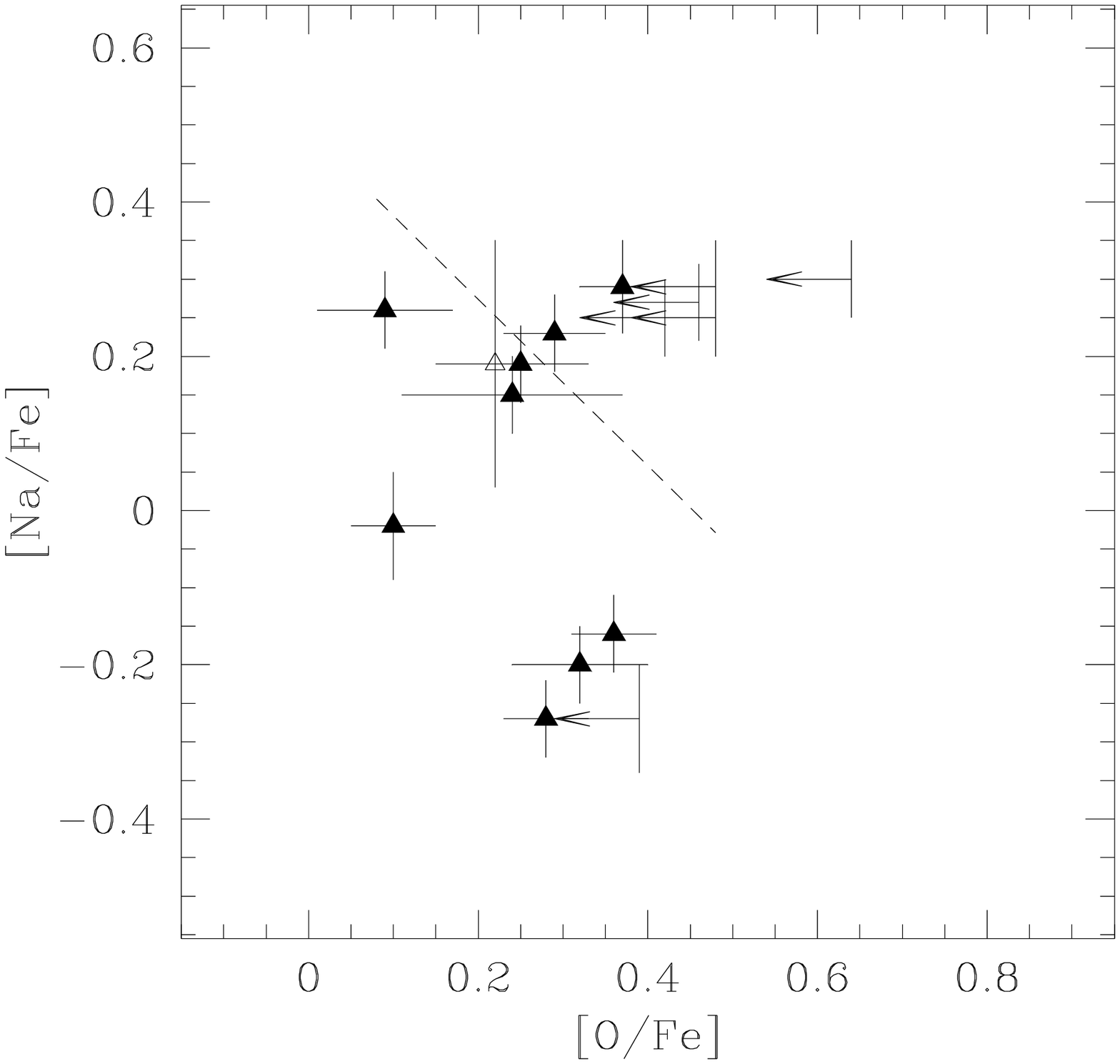}
\figcaption[o_na_m5.ps]{[Na/Fe] against [O/Fe] for our sample of M5 stars. 
Arrows represent upper limits for the [O/Fe] abundance ratio.
The open triangle corresponds to the mean abundance of the six main
sequence stars
The dashed line corresponds to the Na--O anti-correlation present in 
M4 from the analysis of \citet{iva99}, shown as a fiducial line.
\label{o_na_m5}}
\end{figure}

To explore the presence of an anti-correlation between [Na/Fe] 
and [O/Fe] within our M5 sample, we construct Figure~\ref{o_na_m5},
which presents the Na versus O abundance diagram for 15 stars
of our sample in M5 where both sodium and oxygen abundance ratios
were derived.
Our data are indicated by filled triangles and the
arrows represent the upper limits for [O/Fe].
The open triangle represent the mean value for the main sequence
stars, whose [O/Fe] is obtain through the equivalent widths measured
in the summed spectra and whose [Na/Fe] is the average of the
values of the main sequence stars derived individually.
In view of the large sample of bright RGB stars studied in M4 by
\citet{iva99}, we adopt their results for the observed anti-correlation
between Na and O among red giants in this cluster to provide a
fiducial line for a visual comparison.
The anti-correlation found from their sample is indicated as a dashed line
in Figure~\ref{o_na_m5}.
We find no statistically significant anti-correlation between 
[Na/Fe] and [O/Fe] among the nine stars in our sample of M5 with both
clear detections of Na I and O I lines. Nevertheless, we cannot rule 
out its existence when considering that such an anti-correlation might 
be hidden among the stars with only upper limits for their oxygen abundance 
ratio.  

\begin{figure}
\epsscale{0.7}
% Comment out the following line to embed the PS figure into the manuscript
\plotone{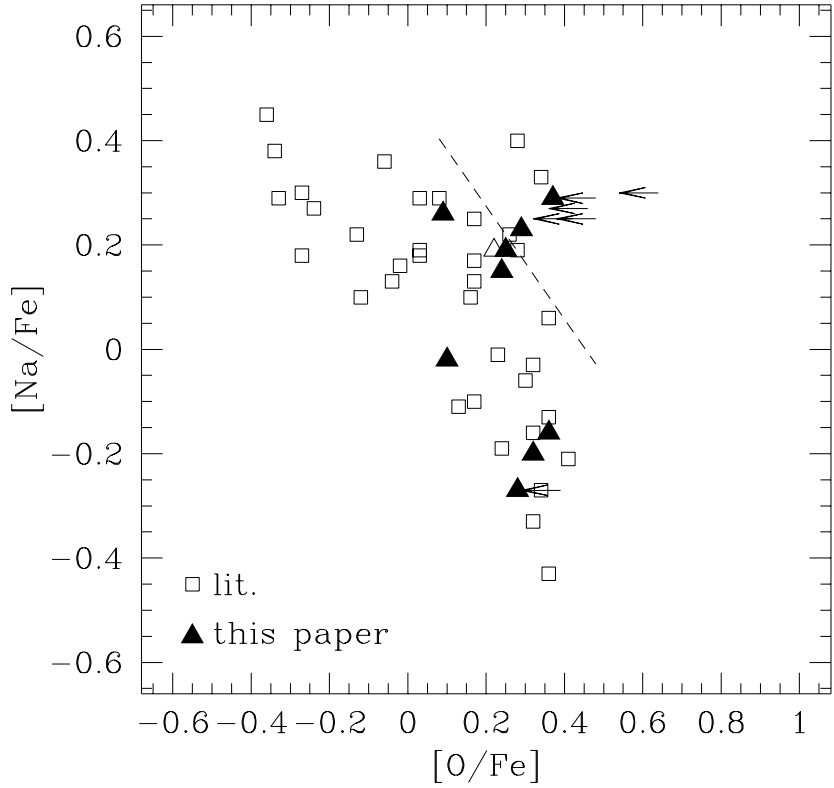}
\figcaption[o_na_lit.ps]{[Na/Fe] against [O/Fe] for
stars in M5 from our analysis (filled triangles for clear detections, 
open triangle for mean main sequence stars) and others from the 
literature \citep{iva01,she96,sne92} (open squares).
Arrows represent our upper limits for the [O/Fe] abundance ratio.
The dashed line corresponds to the anti-correlation observed in M4 
from \citet{iva99}, shown as a fiducial line.
\label{o_na_lit}}
\end{figure}

In Figure~\ref{o_na_lit}, we compare our determination of Na and O 
abundance ratios, in triangles (filled triangles for the M5
RGB and SGB stars and an
open triangle for the mean of the main sequence stars), 
with those from 36 giant stars in M5
from \cite{iva01}, \cite{she96} and \cite{sne92}, 
depicted as open squares.
The dashed line corresponds to the anti-correlation observed in M4
from \citet{iva99}, adopted here and in \cite{ram02} as a fiducial line.
Our observed range in [Na/Fe] is similar to the range observed by these
other studies, but our range in [O/Fe] 
is considerably smaller. Note that
$\sim$1/3 of the stars in the sample of \cite{iva99} have
[O/Fe]$ \le 0.0$ dex, while no star in our sample has
such a low O,
probably because such stars in the luminosity range of our sample
would have O I lines that are not detectable
at our spectral resolution and signal-to-noise ratio.

\subsection{Comparison with \cite{iva01} \label{ivans_comp} }

We summarize here the comparison of our results with those of \cite{iva01},
which is the most extensive previous high dispersion abundance analysis 
for M5.  Recall that \cite{iva01} considered only luminous
red giants and AGB stars.   There is only one star in common
between the two samples, M5 IV-81.  We assign to this star
a \teff\ 85 K higher than the value assigned by \cite{iva01}.
Extracting $V-I$ colors from the database of \cite{ste00} for
some of the stars in the sample of Ivans \etal\
and then determining \teff\ with these colors exactly
as was done for our sample of M5 stars, we find that
at a given $V$ mag along the RGB, the \teff\ adopted here
are about 90 K hotter for stars cooler than 4300 K, with
smaller differences for somewhat hotter stars, to the hottest
stars in their sample, with \teff\ $\sim4700$ K.  For a given \teff,
we have the same values for \grav.  In view of this close
agreement between the stellar parameters adopted for these
two studies, we compare the two sets of derived abundances 
without any adjustments.

The formal statistical uncertainty in the mean abundance
of an element in a globular cluster determined
by an analysis such as ours is artificially small ($<0.05$ dex) as
there are many stars in each study, often with many detected
lines per star.  These small errors are representative of
internal errors only, or of comparisons between analyses
of different clusters carried out with exactly identical 
procedures and sources and quality of all data, 
including both spectroscopic
and photometric.  However, in comparing between analyses,
it is the systematics of small differences in the temperature
scale, the adopted reddening, the adopted atomic data, etc
which can introduce much larger differences in the derived
abundances.  We therefore
assign to such comparisons an uncertainty given by 
$\sigma_c^2 = \sigma({\rm T_{eff}})^2 +  \sigma({\rm log\  g}))^2 
+  \sigma(\xi)^2 +  \sigma(\fe_{\rm model})^2$, 
i.e. our normal 1$\sigma_{pred}$ uncertainty with the term arising
from errors in \ew\ set to 0.  

Table~\ref{tab_comp} provides a detailed comparison of the               
abundance ratios derived here with those of the study                   
of luminous giants in M5 by \cite{iva01}.  
In the last column of this table we give the value of $\sigma_c$
derived for each element from our analysis.                              
Of the 15 ions in common between our work and that of \cite{iva01},      
the only ones with                                                       
$|$[X/Fe](us)$-$[X/Fe](Ivans)$| > 0.16$ dex  
are  O~I and Ba~II.
The difference of +0.24 dex in the mean O abundance presumably
reflects our inability to detect weak O lines in the                 
O-poor low luminosity part of our sample in M5, assuming they            
are actually present there.

We believe that the difference of +0.24 dex in Ba~II arises from the     
difference in the adopted atomic data, isotopic ratios, 
and Solar abundance between our analysis and that of Ivans \etal;
the contribution of the last (given in Table~\ref{tab_solar}) is 0.06 dex.
The Ba~II absorption lines are quite strong among the more luminous
RGB starsin M5.  They have hyperfine structure,
and are quite sensitive to the choice of
microturbulent velocity (see Table~\ref{tab_sensit}) 
as well as damping constants.

Given that one believes, based on the evidence presented here
and in our earlier work on M71, that  the
abundances of most, but not all, elements in globular cluster are constant
from star to star,
one can use the two recent analyses of stars in M5,
our analysis and that of \cite{iva01}, as an end-to-end test of the
accuracy of abundance analyses. The two are completely independent,
using different model stellar atmospheres grids, different procedures
to assign \teff, different ways to measure \ew\ 
and independent choices of atomic data, although the
same abundance analysis code.
We offer above reasonable explanations to explain the discrepancies
in the case of the two elements most divergent
between the two analyses.  For O~I, the explanation involves
sample selection and is thus outside the realm of the analysis,
while for Ba~II it involves the choice of atomic parameters,
factors intrinsic
to the analysis itself.  For the remaining 12 elements in common,
Table~\ref{tab_comp}
serves as an interesting example of how well abundance analyses
can be carried out on globular cluster stars over a wide range
of luminosities in this era of 10-m telescopes.  We note the excellent
agreement of the Fe abundance, $\Delta$[Fe/H] = 0.06 dex, while
for 11 elements in common $\Delta$[X/H](us - Ivans) = $+0.02 \pm0.10$ dex                                      
   
\subsection{Comparison with M71 \label{section_m71comp}}

\begin{figure}
\epsscale{0.7}
% Comment out the following line to embed the PS figure into the manuscript
\plotone{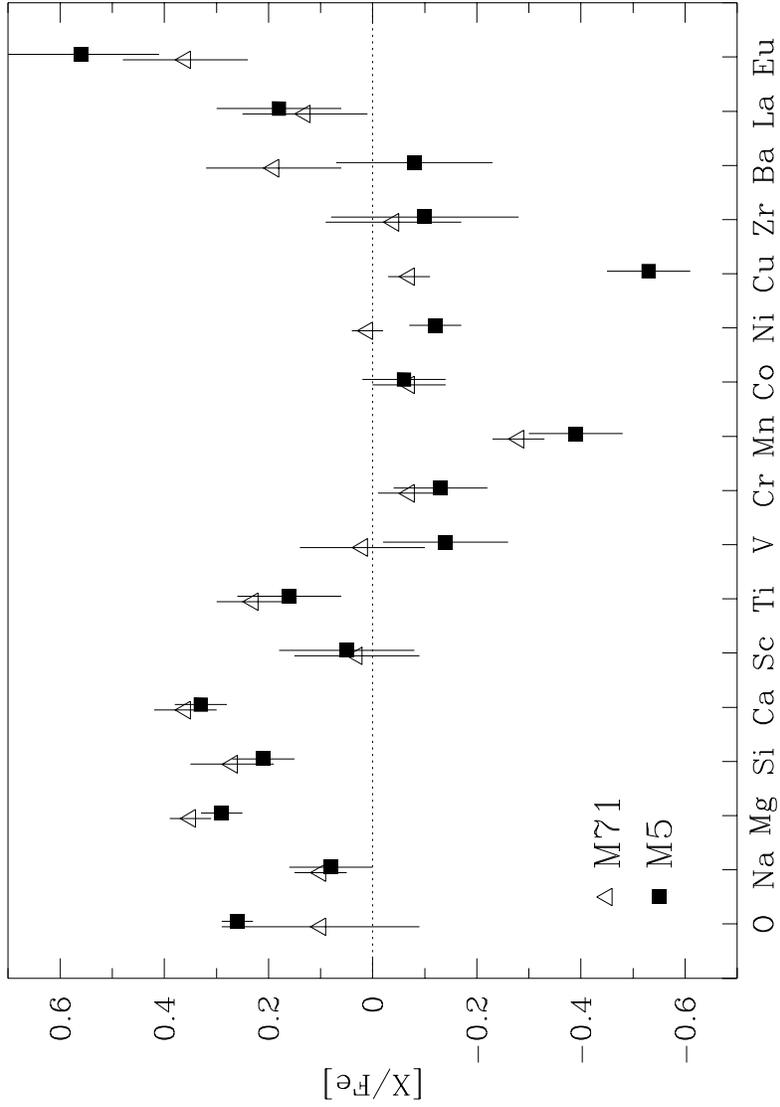}
\figcaption[common.ps]{A comparison of the abundance ratios for
all elements in common between 
our analysis of similar data for 25 stars in 
M71 \citep{ram02}, shown by triangles, and 
in M5 (this paper), shown as squares. The error bars correspond to the
$\sigma_c$ calculated for each cluster.
\label{common}}
\end{figure}

In Figure~\ref{common}, we provide a comparison between our previous 
high-resolution abundance analyses of a comparable sample of stars
in M71 with the present sample in M5
for every element in common to both analyses. 
Our previously published 
M71 abundance ratios have been updated to the same solar abundance scale
adopted here for the M5 abundance  
ratios; see \S\ref{sec_newm71} and Table~\ref{table_newm71}.
The filled squares correspond to our current analysis of M5 stars 
and the filled triangles to our previous analysis of
M71 stars \citep{ram02}. 
The error bars indicate $\sigma_c$ for each abundance ratio.
Overall M5 appears to have abundance ratios very similar to those of  M71;
the mean abundances for each element determined in each globular
cluster agree to within the 1$\sigma_c$ uncertainties of each measurement 
for most of the elements displayed. 
We note that [O/Fe] in M5 appears to be at the top of the range
seen in M71, where \cite{ram01} did detect a definite range, again
suggesting that in M5 we are failing to detect the full range in
O abundance due to weak lines and inadequate precision spectra,
and that at least some of
our non-detections or upper limits do in fact correspond
to low [O/Fe] stars in M5, as were seen by \cite{iva01} among the
stars near the tip of the RGB. 

Nickel and copper show differences in abundance relative to Fe
larger than 2$\sigma_c$.  In each case,
the abundance ratio [X/Fe] is smaller in M5 than it is in M71.
Nickel is discrepant primarily because its predicted
dispersion is so low, 0.05 dex.  This occurs because Ni is so similar to
Fe in its properties that many errors cancel in forming the
abundance ratio [Ni/Fe].  The actual
difference of [Ni/Fe](M5-M71) is only $-0.13$ dex.
The abundance of Cu is based on single line which
has complex hyperfine structure that might not have been
modeled correctly.  

The mean abundances deduced from lines of Ba~II differ
between M71 and M5 by an amount larger
than 0.2 dex, but smaller than 2$\sigma_c$.  
There are three reasonably strong Ba~II
lines used for both the M5 and M71 stars, 
which makes this  perhaps the most 
credible of the proposed differences
between abundance ratios in our sample of stars M71 and in those of M5. 
This might be related to the dependence with metallicity of the production of
$s$- and $r$-process elements \citep{sne01}.  More likely,
however, it too is a reflection of difficulties in  the two analyses.
The Ba absorption lines
tend to be quite strong, hence the choice of damping constants is important.
They also have hyperfine structure whose
pattern depends on the relative abundances of the various
stable isotopes of Ba, which might vary depending on how the Ba is produced.
As shown in
Table~\ref{tab_sensit}, $\sigma_c$ is large
because of the high sensivity of Ba~II lines to changes in $v_t$.
It is interesting to note that the [Ba/Fe] deduced by \cite{iva01} for
the luminous giants in M5 is close to that we derive for M71, where
the Ba~II lines are stronger.

If we were to believe that the Ba abundance difference between M5 and M71 
is real, we should see a similar difference in the La abundance,
as La is also a primarily a s-process element at this metallicity.  
The La abundance,
which is based on three weak lines, is from our work identical in 
M5 and in M71, suggesting that the apparent difference in the Ba abundance
between the two clusters is not real.  

Some of the well known trends characteristic of halo 
star abundances as reviewed by \citet{mcw97}, such as the overabundance
of [$\alpha$/Fe] in metal poor stars, are easily seen in 
Figure~\ref{common}.

\section{Overview and Future Directions}

Now that we have completed a detailed abundance analysis
for a large sample of stars from the RGB tip to near the 
main sequence in two globular clusters, we look back to
summarize the main trends we have seen, offer some brief
remarks in some cases on the possible causes, and some
comments on future direction.

First we focus on the fact that for most elements,
no star to star scatter in abundance in seen which is larger than
that expected from the errors in the observational
data and in the analysis.  For ions with multiple detected lines,
this is a tight constraint, sometimes less than 0.05 dex
and often less than 0.1 dex. 
For this same set of elements, there are no trends
of abundance ratios varying with \teff\ or luminosity
within either of these two globular clusters larger than
$\pm\sim0.1$ dex for the better studied elements.  Furthermore, 
with the minor
caveats noted in \S\ref{section_m71comp}, there are no substantial
changes ($>0.2$ dex) in the abundance ratios between M71 and M5
even though M5 is a factor of 4 more metal poor.

In addition to their implications for the formation of globular
clusters, a subject we defer to the end of this series of papers,
these observational facts provide strong limits for
the potential presence and strength of diffusion processes in the
atmospheres of globular cluster stars,  as discussed in detail
in \S\ref{section_diffusion}.  The results for Fe (with two ions
with multiple detected lines in most stars), and for other
elements as well, constrain the 
amplitude of non-LTE effects (see \S\ref{sec_iron_abun}).

Both diffusion and non-LTE are expected to produce larger
perturbations at lower metallicity.  We now have a well
understood analysis procedure, which we have verified
from end to end through comparison of our results with those of
an independent analysis of much brighter stars in M5 by \cite{iva01}.
We are now ready to move to the extremely metal poor globular
clusters, as we intend to do in the next papers in this series,
with confidence in the validity of our work.

Finally, we turn to the even more interesting case of elements
which do show star to star variations.  We have
seen Na, O and Al vary in a correlated manner in M71.
In M5, where Al was not observed and detecting variations in 
O is quite hard due to the very weak lines in the faint low luminosity
stars in our sample, we detected Na varying.  In neither cluster
did we see Mg varying.  To first order,
these variations have a constant amplitude independent of luminosity.
Much more detailed information  
is available for C and N from study of molecular bands
in the spectra of even larger samples of stars in M5 and in M71
\citep{coh99,bri01a,coh02}.

There are many possible explanations for these variations, which we
may roughly classify into primordial variations, 
internal nucleosynthesis within a star plus mixing to the surface of
that star, and production of material in some other star which somehow
(through stellar winds, supernovae, etc.)
then gets accreted onto the star we actually observe.  (A
variant of the latter scheme involving planets is quite popular
lately for some other problems.)  These theories each predict
the minimum luminosity at which such effects can be detected.
For example, if a star is of such a low mass that Na cannot be produced,
irrespective of whether or not mixing of material to the 
surface is feasible, no Na variations at the stellar surface can be created.
A scenario invoking primordial
variations within the gas cloud
from which the globular cluster formed does not seem to be
able to explain why only certain selected elements are varying.
Detailed theoretical study of the relevant mechanisms includes
meridional mixing \citep{swe79},
turbulent diffusion \citep{cha94,cha95}, proton-capture nuclear
reactions together with deep mixing \citep{lan93,den90,den96,cav98,wei00},
and self pollution \citep{ven01}. 

A key feature is the predicted lowest luminosity (lowest mass) in
an old metal poor population (i.e. a globular cluster) at which
deep mixing can be effective, which usually is at or near that of the
the bump in the luminosity function of the RGB \citep{bon01}.
The location of the bump of the RGB in a cluster of the metallicity
of M5 is predicted to be around 0.2 magnitudes above the horizontal 
branch \citep{zoc99}.  This is
considerably brighter than the stars among which
we found variations in O, Na and Mg
in M71 or the variations in Na detected here in M5.

A second key feature is the amplitude of and
correlation among variations, e.g.
in our dataset we find that, in particular, that in M71
Na and O are anti-correlated, while Na and Al are correlated.
These are signatures to the process by which this material
was synthesized.  We suggest that to first order the
amplitude of the abundance variations appears
to be independent of luminosity, another important clue.

The weight of the evidence is now overwhelming that substantial
variations exist in the elements C, N, O, Na and Al well below
the cutoff luminosity of the RGB bump.  Variations in
all of these elements, except perhaps Al, reach to even lower luminosities,
below the base of the giant branch to 
the main sequence in old metal-poor stellar populations.  
Current theories are hard-pressed 
to reproduce the trends seen among low luminosity stars in globular clusters
in the mounting body of observational
data on abundance variations from our own work here and in M71
\citep{ram02} as well as that of others
(47 Tuc by Cannon \etal\ 1998 and NGC 6752 by Gratton \etal\ 2001).
We can only hope that continued work in this field, particularly
our our upcoming papers on even more metal poor globular clusters, and
continued parallel work on variation of the molecular bands in these
stars, combined with the continued development of various
theoretical aspects of stellar evolution, 
will lead us to new insights and resolution of these issues.

\acknowledgements
The entire Keck/HIRES user communities owes a huge debt to 
Jerry Nelson, Gerry Smith, Steve Vogt, and many other 
people who have worked to make the Keck Telescope and HIRES  
a reality and to operate and maintain the Keck Observatory. 
We are grateful to the W. M.  Keck Foundation for the vision to fund
the construction of the W. M. Keck Observatory. 
The authors wish to extend special thanks to those of Hawaiian ancestry
on whose sacred mountain we are privileged to be guests. 
Without their generous hospitality, none of the observations presented
herein would have been possible.  We acknowledge a special debt to
Brad Behr, who participated in the very early stages of this project.
We thank Peter Stetson for providing his photometry database in digital form.
We are grateful to the National Science Foundation for partial support under
grant AST-9819614 and AST-0205951 to JGC.  We thank Jason Prochaska 
and Andy McWilliam
for providing their tables of hyperfine structure in digital form.

\appendix
\section{Updated M71 Abundances \label{sec_newm71}}

For convenience, we present in Table~\ref{table_newm71} the updated
abundance ratios for M71 after the modifications described here
of updating our adopted Solar abundances as described
in \S\ref{section_sun} and correcting an error in the
transition probability used for the 6497\AA\ line of Ba~II in
our earlier work on M71 \citep{ram02}.  An error in the Ca~I damping
constants used in our earlier work has also been corrected here.

\clearpage

% REFERENCES

%Figuras

\clearpage

% Tablas 

%
% Tabla1
% 
\clearpage
\begin{deluxetable}{lrrrrr}
%\rotate
%\small
%\footnotesize
%\scriptsize
\tablenum{1}
\tablewidth{0pt}
\tablecaption{The Sample of Stars in M5. \label{tab_sample}}
\tablehead{
\colhead{ID\tablenotemark{a}} & 
\colhead{V} &
\colhead{Date Obs.} & 
\colhead{Exp. Time} &
\colhead{SNR\tablenotemark{b}} &
\colhead{$v_{r}$} \\ 
\colhead{} & 
\colhead{(mag)} &   
\colhead{} & 
\colhead{(sec)} &
\colhead{} & \colhead{(\kms)} }
\startdata 
IV-81        & 12.22 & 6/2000 & 700 & $>100$  & +58.1  \\
IV-59        & 12.71 & 6/2000 & 400 & $>100$  & +59.8  \\
IV-82        & 13.24 & 6/2000 & 700 & $>100$  & +48.1  \\
1-109        & 14.25 & 5/2001 & 1400 & $>100$ & +58.9  \\
1-36         & 14.73 & 6/2000 & 7200 & $>100$ & +60.7  \\
1-31         & 14.86 & 6/2000 & 4800 & $>100$ & +57.3  \\
1-40         & 14.92 & 6/2000 & 3600 & $>100$ & +55.1  \\
G18484\_0316 & 15.41 & 6/2000 & 4800 & $>100$ & +55.3  \\
1-110        & 15.43 & 5/2001 & 1400 &     90 & +57.3  \\
G18484\_0309 & 15.93 & 6/2000 & 4800 & $>100$ & +58.1  \\
G18450\_0453 & 16.26 & 6/2000 & 6000 &     75 & +54.2  \\
1-32         & 16.51 & 6/2000 & 4800 &     92 & +57.9  \\
G18458\_0547 & 16.63 & 6/2000 & 7200 &     89 & +50.2  \\
G18447\_0453 & 16.69 & 6/2000 & 6000 &     75 & +49.4  \\
G18155\_0228 & 17.11 & 5/2001 & 7200 &     64 & +52.0  \\
G18564\_0457 & 17.06 & 5/2001 & 7200 &     64 & +53.4  \\
G18445\_0448 & 17.13 & 6/2000 & 6000 &     88 & +63.8  \\
G18483\_0608 & 17.21 & 5/2001 & 7200 &     68 & +53.6  \\
G18279\_0101 & 17.21 & 5/2001 & 7200 &     66 & +53.8  \\
G18569\_0455 & 18.07 & 5/2001 & 7200 &     42 & +50.6  \\ 
G18152\_0232 & 18.09 & 5/2001 & 7200 &     45 & +48.6  \\  
G18487\_0606 & 18.09 & 5/2001 & 7200 &     43 & +54.6  \\
G18172\_0750 & 18.11 & 5/2001 & 7200 &     50 & +59.7  \\
G18279\_0107 & 18.14 & 5/2001 & 7200 &     43 & +47.6  \\
G18175\_0749 & 18.14 & 5/2001 & 7200 &     41 & +56.7  \\
\enddata
\tablenotetext{a}{Identifications are from \citet{arp62}, from \cite{buo81}
or are assigned based on the J2000 coordinates, 15 rm rs.s +2 dm dd becoming
Grmrss\_dmdd.}
\tablenotetext{b}{Signal to noise ratio in the continuum near
6150 \AA\ per 4 pixel spectral resolution element.}
\end{deluxetable}

%
% Tabla2
%
\clearpage                                                                      
\begin{deluxetable}{llrrcccccccccc}                                          
\rotate                                                                         
\small                                                                         
\footnotesize                                                                  
\scriptsize                                                                    
\tablenum{2}                                                                    
\tablewidth{0pt}                                                                
\tablecaption{Equivalent Widths (m$\AA$)\tablenotemark{a}}                      
\tablehead{                                                                     
\colhead{Ion} & \colhead{$\lambda (\AA)$} & \colhead{$\chi$ (eV)} & \colhead{log ($gf$)} & \colhead{IV--81} & \colhead{IV--59} & \colhead{IV-82} & \colhead{1--109} & \colhead{1--36} & \colhead{1--31} & \colhead{1--40} & \colhead{G18481} & \colhead{1--110} & \colhead{G18484} \\                                                                                                                                                                                                                                                       
\colhead{} & \colhead{} & \colhead{} & \colhead{} & \colhead{} & \colhead{} & \colhead{} & \colhead{} & \colhead{} & \colhead{} & \colhead{} & \colhead{\_0316} & \colhead{} & \colhead{\_0309} }                                                                                                                                                                                                                                                                                                                                                
\startdata                                                                      
O I   & 6300.304 &   0.000 &  -9.780 &    60.0 \tablenotemark{b}  &    70.0 \tablenotemark{b}  &    45.0 \tablenotemark{b}  & \nodata                    &    20.0 \tablenotemark{b}  &    16.0 \tablenotemark{b}  &    14.0 \tablenotemark{b}  &    16.0 \tablenotemark{b}  & \nodata                    & \nodata                    \\
O I   & 6363.776 &   0.020 & -10.300 &    29.0 \tablenotemark{b}  &    19.0 \tablenotemark{b}  &    14.0 \tablenotemark{b}  & \nodata                    & \nodata                    & \nodata                    & \nodata                    & \nodata                    & \nodata                    & \nodata                     \\
O I   & 7771.944 &   9.150 &   0.369 & \nodata                    & \nodata                    &    14.0 \tablenotemark{b}  &    11.0 \tablenotemark{b}  &    17.5 \tablenotemark{b}  &    25.0 \tablenotemark{b}  &    22.0 \tablenotemark{b}  &    27.0 \tablenotemark{b}  & \nodata                    &    25.0 \tablenotemark{b}   \\
O I   & 7774.166 &   9.150 &   0.223 & \nodata                    & \nodata                    &    12.0 \tablenotemark{b}  &     9.0 \tablenotemark{b}  &    26.5 \tablenotemark{b}  &    29.0 \tablenotemark{b}  &    19.0 \tablenotemark{b}  &    24.0 \tablenotemark{b}  &    21.5 \tablenotemark{b}  &    18.0 \tablenotemark{b}   \\
O I   & 7775.388 &   9.150 &   0.001 & \nodata                    & \nodata                    & \nodata                    &    10.0 \tablenotemark{b}  &     6.0 \tablenotemark{b}  &    12.0 \tablenotemark{b}  &    10.0 \tablenotemark{b}  & \nodata                    & \nodata                    &    15.0 \tablenotemark{b}   \\
Na I  & 5682.633 &   2.100 &  -0.700 &   102.5 \tablenotemark{b}  &    71.8 \tablenotemark{b}  &    49.7 \tablenotemark{b}  &    57.1 \tablenotemark{b}  &    45.4 \tablenotemark{b}  &    31.8 \tablenotemark{b}  &    50.4 \tablenotemark{b}  &    16.9 \tablenotemark{b}  &    42.4 \tablenotemark{b}  &    19.3 \tablenotemark{b}   \\
Na I  & 5688.193 &   2.100 &  -0.420 &   121.5 \tablenotemark{b}  &   101.3 \tablenotemark{b}  &    69.2 \tablenotemark{b}  &    78.6 \tablenotemark{b}  &    65.4 \tablenotemark{b}  &    45.9 \tablenotemark{b}  &    70.0 \tablenotemark{b}  &    42.2 \tablenotemark{b}  &    66.4 \tablenotemark{b}  &    31.8 \tablenotemark{b}   \\
Na I  & 5889.950 &   0.000 &   0.110 &   700.0 \tablenotemark{b}  &   470.8 \tablenotemark{b}  &   322.3 \tablenotemark{b}  &   347.0 \tablenotemark{b}  &   309.0 \tablenotemark{b}  &   269.1 \tablenotemark{b}  &   327.2 \tablenotemark{b}  &   257.2 \tablenotemark{b}  &   306.2 \tablenotemark{b}  &   232.5 \tablenotemark{b}   \\
Na I  & 5895.924 &   0.000 &  -0.190 &   539.0 \tablenotemark{b}  &   382.5 \tablenotemark{b}  &   276.3 \tablenotemark{b}  &   285.4 \tablenotemark{b}  &   258.4 \tablenotemark{b}  &   227.1 \tablenotemark{b}  &   271.1 \tablenotemark{b}  &   217.4 \tablenotemark{b}  &   270.7 \tablenotemark{b}  &   202.4 \tablenotemark{b}   \\
Na I  & 6154.225 &   2.100 &  -1.530 &    36.4 \tablenotemark{b}  &    22.3 \tablenotemark{b}  &     6.0 \tablenotemark{b}  &    19.4 \tablenotemark{b}  &     9.0 \tablenotemark{b}  &     6.0 \tablenotemark{b}  &     7.6 \tablenotemark{b}  & \nodata                    & \nodata                    & \nodata                     \\
Na I  & 6160.747 &   2.100 &  -1.230 &    58.7 \tablenotemark{b}  &    35.5 \tablenotemark{b}  &    14.2 \tablenotemark{b}  &    29.6 \tablenotemark{b}  &    17.4 \tablenotemark{b}  &     8.5 \tablenotemark{b}  &    19.9 \tablenotemark{b}  &     6.5 \tablenotemark{b}  &    23.4 \tablenotemark{b}  & \nodata                     \\
Mg I  & 5528.405 &   4.340 &  -0.480 &   183.8                    &   188.3                    &   157.6                    & \nodata                    &   131.6                    &   139.7                    &   145.3                    &   128.6                    & \nodata                    &   109.1                     \\
Si I  & 5421.178 &   5.620 &  -1.430 &    15.8                    & \nodata                    & \nodata                    & \nodata                    &    11.1                    & \nodata                    & \nodata                    & \nodata                    & \nodata                    & \nodata                     \\
Si I  & 5665.554 &   4.920 &  -2.040 &    24.8                    & \nodata                    &    19.6                    & \nodata                    &    15.6                    &    13.5                    & \nodata                    &    13.1                    & \nodata                    & \nodata                     \\
Si I  & 5690.427 &   4.930 &  -1.870 &    28.6                    &    28.6                    &    25.4                    &    28.7                    &    21.6                    &    24.3                    &    21.3                    &    18.0                    & \nodata                    & \nodata                     \\
Si I  & 5701.105 &   4.930 &  -2.050 & \nodata                    & \nodata                    &    20.8                    & \nodata                    &    15.3                    &    15.6                    & \nodata                    &    12.6                    & \nodata                    &    16.7                     \\
Si I  & 5772.145 &   5.080 &  -1.750 &    21.2                    &    28.4                    &    27.1                    &    22.9                    &    19.5                    &    18.9                    &    18.7                    &    15.0                    & \nodata                    &    16.7                     \\
Si I  & 5793.071 &   4.930 &  -2.060 &    23.8                    &    24.9                    &    20.8                    & \nodata                    &    17.1                    & \nodata                    &    15.6                    & \nodata                    & \nodata                    & \nodata                     \\
Si I  & 5948.540 &   5.080 &  -1.230 &    54.0                    &    61.6                    &    60.5                    &    55.3                    &    48.7                    &    43.4                    &    43.7                    &    36.7                    & \nodata                    &    23.4                     \\
Si I  & 6145.015 &   5.610 &  -1.440 &    12.8                    & \nodata                    &    15.0                    & \nodata                    &    13.8                    & \nodata                    &    10.3                    &     8.6                    & \nodata                    & \nodata                     \\
\enddata                                                                        
\tablenotetext{a}{Table available electronically.}                              
\tablenotetext{b}{Line identified by hand. All other lines are                  
identified automatically.}                                                      
\tablenotetext{c}{Fe I line used in the $\lambda D-W_{\lambda} $ fit.}          
\end{deluxetable}                                                               
                                                                                
%
% Tabla3
%
\clearpage
\begin{deluxetable}{lcccc}
%\rotate
%\small
%\footnotesize
%\scriptsize
\tablenum{3}
\tablewidth{0pt}
\tablecaption{Solar Abundance Ratios [X/Fe]
\label{tab_solar}}
\tablehead{\colhead{Ion} & \colhead{\# lines} &
\colhead{[X/Fe]\tablenotemark{a}} &
\colhead{$\sigma$\tablenotemark{a}}
& \colhead{$\Delta$[us-photospheric]\tablenotemark{b}} \\
\colhead{} & \colhead{} & \colhead{(dex)} & \colhead{(dex)} & \colhead{(dex)}}
\startdata
O I   &  2 &  +1.60 & 0.08 &  +0.27 \\
Na I  &  4 & --1.21 & 0.04 & --0.04 \\
Mg I  &  9 &  +0.02 & 0.18 & --0.06 \\
Si I  & 26 &  +0.09 & 0.11 &  +0.04 \\
Ca I  & 19 & --1.30 & 0.18 & --0.16 \\
Sc II &  7 & --4.24 & 0.12 &  +0.09 \\
Ti I  & 42 & --2.52 & 0.13 & --0.04 \\
V I   & 13 & --3.48 & 0.16 &  +0.02 \\
Cr I  & 13 & --1.78 & 0.12 &  +0.05 \\
Mn I  &  5 & --2.08 & 0.08 &  +0.03 \\
Fe I \tablenotemark{c} &316 &   7.44 & 0.16 & --0.06 \\
Fe II\tablenotemark{c} & 15 &   7.47 & 0.08 & --0.03 \\
Co I  &  9 & --2.49 & 0.16 &  +0.09 \\
Ni I  & 65 & --1.20 & 0.18 &  +0.05 \\
Cu I  &  1 & --3.30 & ...  & --0.01 \\
Zn I  &  1 & --2.88 & ...  &  +0.02 \\
Zr I  &  3 & --4.61 & 0.09 &  +0.29 \\
Ba II &  3 & --5.31 & 0.06 &  +0.06 \\
La II &  1 & --6.30 & ...  &  +0.03\tablenotemark{d} \\
Eu II &  1 & --6.92 & ...  &  +0.01\tablenotemark{e} \\
\enddata
\tablenotetext{a}{Mean and 1$\sigma$ rms deviation about the mean
for the abundance in the Sun of the lines
of a particular ion using
our adopted atomic line parameters.}
\tablenotetext{b}{Photospheric solar abundances from \citet{gre98}.}
\tablenotetext{c}{log $\epsilon$(Fe)}
\tablenotetext{d}{We adopt the Solar La abundance of \cite{lawler01a} 
($\epsilon(La)=+1.14$ dex), which is 0.03 dex smaller than
that of \cite{gre98}.  See the discussion in \S\ref{section_sun}.}
\tablenotetext{d}{We adopt the Solar Eu abundance of \cite{lawler01b}
($\epsilon(Eu)=+0.52$ dex), which is 0.01 dex higher
than that of \cite{gre98}.  See the discussion in \S\ref{section_sun}.}
\end{deluxetable}

%
% Tabla4
%
\clearpage
\begin{deluxetable}{lccc}
%\rotate
%\small
%\footnotesize
%\scriptsize
\tablenum{4}
\tablewidth{0pt}
\tablecaption{Stellar Parameters for the M5 Sample. 
\label{tab_ste}}
\tablehead{
\colhead{ID\tablenotemark{a}} &
\colhead{\teff } &
\colhead{\grav } &
\colhead{\mtv } \\
\colhead{} &
\colhead{(K)} &
\colhead{(dex)} &
\colhead{(km/s)} 
}
\startdata
IV-81        & 4035 & 0.60 & 2.06 \\
IV-59        & 4265 & 1.00 & 1.94 \\
IV-82        & 4375 & 1.20 & 1.89 \\
1-109        & 4635 & 1.80 & 1.76 \\
1-36         & 4755 & 2.10 & 1.70 \\
1-31         & 4880 & 2.25 & 1.64 \\
1-40         & 4810 & 2.20 & 1.67 \\
G18484\_0316 & 4995 & 2.50 & 1.58 \\
1-110        & 4875 & 2.50 & 1.64 \\
G18484\_0309 & 5325 & 2.80 & 1.41 \\
G18450\_0453 & 5170 & 2.90 & 1.49 \\
1-32         & 5066 & 3.00 & 1.54 \\
G18458\_0547 & 5090 & 3.00 & 1.53 \\
G18447\_0453 & 5275 & 3.15 & 1.44 \\
G18155\_0228 & 5270 & 3.25 & 1.44 \\
G18564\_0457 & 5400 & 3.40 & 1.37 \\
G18445\_0448 & 5300 & 3.30 & 1.42 \\
G18483\_0608 & 5300 & 3.30 & 1.42 \\
G18279\_0101 & 5335 & 3.35 & 1.41 \\
G18569\_0455 & 5825 & 3.85 & 1.16 \\
G18152\_0232 & 6000 & 3.95 & 1.07 \\
G18487\_0606 & 5930 & 3.90 & 1.11 \\
G18172\_0750 & 6100 & 4.00 & 1.02 \\
G18279\_0107 & 5785 & 3.90 & 1.18 \\
G18175\_0749 & 6000 & 3.95 & 1.07 \\
\enddata
\tablenotetext{a}{Identifications are from \citet{arp62,buo81}
or are assigned based on the J2000 coordinates, rh rm rs.s dd dm dd becoming
Grmrss\_dmdd.}
\end{deluxetable}

%
% Tabla5
%
\begin{deluxetable}{lcccccccccc}
\rotate                         
\small                         
\footnotesize                  
\scriptsize                    
\tablenum{5a}                    
\tablewidth{0pt}                
\tablecaption{Iron Abundance and Abundance Ratios: Na -- Mg
\label{tab_ratio_a}}
\tablehead{                     
\colhead{Star} &                
\colhead{${\rm N_{FeI }}$} & \colhead{[Fe/H]$_{I }$} &
\colhead{${\rm N_{FeII}}$} & \colhead{[Fe/H]$_{II}$} &
\colhead{${\rm N_{O }}$} & \colhead{[O /Fe]} &
\colhead{${\rm N_{Na}}$} & \colhead{[Na/Fe]} &
\colhead{${\rm N_{Mg}}$} & \colhead{[Mg/Fe]} }
\startdata
IV--81       & 201 & --1.28$\pm$0.03  &   9 & --1.41$\pm$0.05  &   2 &   0.37$\pm$0.05  &   6 &   0.29$\pm$0.06  &   2 &   0.30$\pm$0.12 \\
IV--59       & 168 & --1.32$\pm$0.03  &   9 & --1.27$\pm$0.05  &   2 &   0.25$\pm$0.08  &   6 &   0.19$\pm$0.05  &   2 &   0.41$\pm$0.05 \\
IV--82       & 187 & --1.35$\pm$0.03  &   9 & --1.36$\pm$0.05  &   4 &   0.28$\pm$0.05  &   6 & --0.27$\pm$0.05  &   2 &   0.23$\pm$0.08 \\
1--109       & 110 & --1.35$\pm$0.03  &   3 & --1.32$\pm$0.05  &   3 &   0.09$\pm$0.08  &   6 &   0.26$\pm$0.05  &   0 &   \nodata       \\
1--36        & 160 & --1.39$\pm$0.03  &   9 & --1.29$\pm$0.05  &   4 &   0.24$\pm$0.13  &   6 &   0.15$\pm$0.05  &   2 &   0.22$\pm$0.10 \\
1--31        & 154 & --1.22$\pm$0.03  &   9 & --1.25$\pm$0.05  &   4 &   0.32$\pm$0.08  &   6 & --0.20$\pm$0.05  &   2 &   0.28$\pm$0.08 \\
1--40        & 131 & --1.36$\pm$0.03  &   5 & --1.33$\pm$0.05  &   4 &   0.29$\pm$0.06  &   6 &   0.23$\pm$0.05  &   2 &   0.34$\pm$0.05 \\
G18484\_0316 & 109 & --1.30$\pm$0.03  &   6 & --1.27$\pm$0.05  &   3 &   0.36$\pm$0.05  &   5 & --0.16$\pm$0.05  &   2 &   0.26$\pm$0.06 \\
1--110       &  73 & --1.38$\pm$0.03  &   4 & --1.21$\pm$0.05  &   1 &  $<$0.42  &   5 &   0.25$\pm$0.05  &   0 &   \nodata       \\
G18484\_0309 &  83 & --1.18$\pm$0.03  &   6 & --1.32$\pm$0.09  &   3 &   0.10$\pm$0.05  &   4 & --0.02$\pm$0.07  &   2 &   0.15$\pm$0.10 \\
G18450\_0453 &  80 & --1.31$\pm$0.03  &   3 & --1.28$\pm$0.05  &   1 &  $<$0.64  &   5 &   0.30$\pm$0.05  &   1 &   0.19$\pm$0.10 \\
1--32        &  75 & --1.45$\pm$0.03  &   2 & --1.30$\pm$0.05  &   1 &  $<$0.48  &   5 &   0.29$\pm$0.06  &   2 &   0.22$\pm$0.30 \\
G18458\_0547 &  68 & --1.44$\pm$0.03  &   5 & --1.13$\pm$0.12  &   0 &   \nodata        &   5 &   0.02$\pm$0.07  &   1 &   0.16$\pm$0.10 \\
G18447\_0453 &  52 & --1.29$\pm$0.05  &   1 & --1.24$\pm$0.05  &   0 &   \nodata        &   4 &   0.18$\pm$0.06  &   2 &   0.17$\pm$0.06 \\
G18144\_0228 &  47 & --1.23$\pm$0.04  &   1 & --1.27$\pm$0.05  &   0 &   \nodata        &   3 & --0.26$\pm$0.05  &   2 &   0.32$\pm$0.06 \\
G18564\_0457 &  59 & --1.01$\pm$0.03  &   0 &   \nodata        &   1 &  $<$0.39  &   4 & --0.27$\pm$0.07  &   1 &   0.19$\pm$0.10 \\
G18445\_0448 &  60 & --1.27$\pm$0.03  &   4 & --1.09$\pm$0.05  &   1 &  $<$0.46  &   4 &   0.27$\pm$0.05  &   1 &   0.42$\pm$0.10 \\
G18483\_0608 &  51 & --1.24$\pm$0.03  &   0 &   \nodata        &   1 &  $<$0.48  &   4 &   0.21$\pm$0.05  &   2 &   0.38$\pm$0.07 \\
G18279\_0101 &  48 & --1.18$\pm$0.07  &   1 & --1.34$\pm$0.05  &   0 &   \nodata        &   4 &   0.15$\pm$0.05  &   0 &   \nodata       \\
G18579\_0455 &  10 & --1.17$\pm$0.08  &   0 &   \nodata        &   0 &   \nodata        &   3 &   0.17$\pm$0.10  &   0 &   \nodata       \\
G18152\_0232 &   6 & --1.13$\pm$0.14  &   0 &   \nodata        &   0 &   \nodata        &   2 & --0.25$\pm$0.05  &   0 &   \nodata       \\
G18487\_0606 &   6 & --1.43$\pm$0.08  &   0 &   \nodata        &   0 &   \nodata        &   2 &   0.74$\pm$0.18  &   1 &   0.39$\pm$0.10 \\
G18172\_0750 &   3 & --1.00$\pm$0.11  &   0 &   \nodata        &   0 &   \nodata        &   2 & --0.26$\pm$0.12  &   0 &   \nodata       \\
G18279\_0107 &   7 & --1.48$\pm$0.12  &   0 &   \nodata        &   0 &   \nodata        &   2 &   0.33$\pm$0.12  &   0 &   \nodata       \\
G18175\_0749 &   5 & --1.22$\pm$0.10  &   0 &   \nodata        &   0 &   \nodata        &   2 &   0.40$\pm$0.25  &   0 &   \nodata       \\
$<$MS$>$     &   0 &   \nodata        &   0 &   \nodata        &   3 &   0.22$\pm$0.07  &   0 &   \nodata        &   1 &   0.11$\pm$0.10 \\
\enddata                                          
\end{deluxetable}                                 
\clearpage                      
 
\begin{deluxetable}{lcccccccccc}
\rotate                         
\small                         
\footnotesize                  
\scriptsize                    
\tablenum{5b}                    
\tablewidth{0pt}                
\tablecaption{Abundance Ratios: Si -- V
\label{tab_ratio_b}} 
\tablehead{                     
\colhead{Star} &                
\colhead{${\rm N_{Si}}$} & \colhead{[Si/Fe]} &
\colhead{${\rm N_{Ca}}$} & \colhead{[Ca/Fe]} &
\colhead{${\rm N_{Sc}}$} & \colhead{[Sc/Fe]} &
\colhead{${\rm N_{Ti}}$} & \colhead{[Ti/Fe]} &
\colhead{${\rm N_{V }}$} & \colhead{[V /Fe]} }
\startdata
IV--81       &  16 &   0.31$\pm$0.06  &  19 &   0.28$\pm$0.06  &   7 &   0.15$\pm$0.08  &  35 &   0.35$\pm$0.04  &   9 &   0.01$\pm$0.05 \\
IV--59       &  12 &   0.24$\pm$0.07  &  17 &   0.25$\pm$0.07  &   7 &   0.18$\pm$0.06  &  28 &   0.08$\pm$0.03  &   9 & --0.25$\pm$0.06 \\
IV--82       &  15 &   0.30$\pm$0.05  &  17 &   0.33$\pm$0.07  &   7 &   0.09$\pm$0.07  &  27 &   0.17$\pm$0.03  &  10 & --0.08$\pm$0.05 \\
1--109       &   9 &   0.27$\pm$0.06  &  15 &   0.26$\pm$0.05  &   5 &   0.13$\pm$0.06  &  10 &   0.09$\pm$0.05  &   4 & --0.27$\pm$0.06 \\
1--36        &  16 &   0.23$\pm$0.05  &  17 &   0.35$\pm$0.05  &   7 &   0.15$\pm$0.05  &  14 &   0.15$\pm$0.03  &   4 & --0.14$\pm$0.05 \\
1--31        &  14 &   0.20$\pm$0.06  &  18 &   0.32$\pm$0.05  &   7 &   0.01$\pm$0.05  &  17 &   0.12$\pm$0.03  &   7 & --0.16$\pm$0.05 \\
1--40        &  12 &   0.25$\pm$0.06  &  18 &   0.37$\pm$0.06  &   7 &   0.07$\pm$0.07  &  14 &   0.14$\pm$0.03  &   3 & --0.17$\pm$0.05 \\
G18484\_0316 &  14 &   0.19$\pm$0.05  &  18 &   0.40$\pm$0.06  &   7 &   0.03$\pm$0.05  &   9 &   0.14$\pm$0.05  &   2 & --0.01$\pm$0.10 \\
1--110       &   5 &   0.15$\pm$0.06  &  16 &   0.31$\pm$0.05  &   6 &   0.22$\pm$0.09  &   4 &   0.12$\pm$0.05  &   1 & --0.40$\pm$0.10 \\
G18484\_0309 &   9 &   0.28$\pm$0.11  &  18 &   0.31$\pm$0.05  &   5 & --0.15$\pm$0.07  &   5 &   0.21$\pm$0.05  &   0 &   \nodata       \\
G18450\_0453 &   6 &   0.19$\pm$0.07  &  17 &   0.40$\pm$0.07  &   5 & --0.04$\pm$0.07  &   6 &   0.24$\pm$0.05  &   0 &   \nodata       \\
1--32        &   4 &   0.12$\pm$0.08  &  18 &   0.40$\pm$0.05  &   5 &   0.15$\pm$0.05  &   4 &   0.20$\pm$0.05  &   1 &   0.01$\pm$0.10 \\
G18458\_0547 &   3 &   0.15$\pm$0.12  &  15 &   0.42$\pm$0.06  &   4 &   0.11$\pm$0.05  &   2 &   0.13$\pm$0.07  &   1 & --0.15$\pm$0.10 \\
G18447\_0453 &   3 &   0.19$\pm$0.11  &  11 &   0.41$\pm$0.07  &   2 & --0.28$\pm$0.13  &   1 &   0.34$\pm$0.10  &   0 &   \nodata       \\
G18144\_0228 &   6 &   0.22$\pm$0.10  &  12 &   0.28$\pm$0.06  &   3 & --0.08$\pm$0.05  &   1 &   0.31$\pm$0.10  &   0 &   \nodata       \\
G18564\_0457 &   5 &  99.00$\pm$0.00  &  15 &   0.26$\pm$0.06  &   1 & --0.12$\pm$0.10  &   4 &   0.15$\pm$0.11  &   0 &   \nodata       \\
G18445\_0448 &   6 &   0.17$\pm$0.08  &  18 &   0.43$\pm$0.05  &   4 &   0.02$\pm$0.05  &   2 &   0.24$\pm$0.13  &   0 &   \nodata       \\
G18483\_0608 &   3 &  99.00$\pm$0.00  &  14 &   0.33$\pm$0.05  &   2 & --0.03$\pm$0.05  &   4 &   0.14$\pm$0.09  &   0 &   \nodata       \\
G18279\_0101 &   2 &   0.06$\pm$0.07  &  12 &   0.24$\pm$0.09  &   2 &   0.03$\pm$0.20  &   3 & --0.06$\pm$0.11  &   0 &   \nodata       \\
G18579\_0455 &   0 &   \nodata        &   2 &   0.29$\pm$0.22  &   0 &   \nodata        &   0 &   \nodata        &   0 &   \nodata       \\
G18152\_0232 &   0 &   \nodata        &   2 & --0.17$\pm$0.23  &   0 &   \nodata        &   0 &   \nodata        &   0 &   \nodata       \\
G18487\_0606 &   0 &   \nodata        &   2 &   0.39$\pm$0.30  &   0 &   \nodata        &   0 &   \nodata        &   0 &   \nodata       \\
G18172\_0750 &   0 &   \nodata        &   3 & --0.05$\pm$0.17  &   0 &   \nodata        &   0 &   \nodata        &   0 &   \nodata       \\
G18279\_0107 &   0 &   \nodata        &   3 &   0.12$\pm$0.16  &   0 &   \nodata        &   0 &   \nodata        &   0 &   \nodata       \\
G18175\_0749 &   0 &   \nodata        &   2 &   0.17$\pm$0.15  &   0 &   \nodata        &   0 &   \nodata        &   0 &   \nodata       \\
$<$MS$>$     &   0 &   \nodata        &   0 &   \nodata        &   2 & --0.27$\pm$0.08  &   0 &   \nodata        &   0 &   \nodata       \\
\enddata                                          
\end{deluxetable}                                 
\clearpage                      
 
\begin{deluxetable}{lcccccccccc}
\rotate                         
\small                         
\footnotesize                  
\scriptsize                    
\tablenum{5c}                    
\tablewidth{0pt}                
\tablecaption{Abundance Ratios: Cr -- Cu
\label{tab_ratio_c}} 
\tablehead{                     
\colhead{Star} &                
\colhead{${\rm N_{Cr}}$} & \colhead{[Cr/Fe]} &
\colhead{${\rm N_{Mn}}$} & \colhead{[Mn/Fe]} &
\colhead{${\rm N_{Co}}$} & \colhead{[Co/Fe]} &
\colhead{${\rm N_{Ni}}$} & \colhead{[Ni/Fe]} &
\colhead{${\rm N_{Cu}}$} & \colhead{[Cu/Fe]} }
\startdata
IV--81       &   9 &   0.02$\pm$0.07  &   3 & --0.16$\pm$0.10  &   6 &   0.06$\pm$0.10  &  46 & --0.08$\pm$0.04  &   1 & --0.45$\pm$0.10 \\
IV--59       &   4 & --0.24$\pm$0.15  &   3 & --0.49$\pm$0.06  &   5 & --0.10$\pm$0.10  &  36 & --0.02$\pm$0.05  &   1 & --0.51$\pm$0.10 \\
IV--82       &   7 &   0.04$\pm$0.09  &   3 & --0.40$\pm$0.07  &   5 & --0.08$\pm$0.08  &  40 & --0.10$\pm$0.04  &   1 & --0.50$\pm$0.10 \\
1--109       &   1 & --0.32$\pm$0.10  &   0 &   \nodata        &   2 &   0.03$\pm$0.20  &  24 & --0.21$\pm$0.05  &   0 &   \nodata       \\
1--36        &   3 & --0.01$\pm$0.20  &   3 & --0.32$\pm$0.07  &   2 &   0.03$\pm$0.11  &  32 & --0.17$\pm$0.05  &   1 & --0.54$\pm$0.10 \\
1--31        &   4 & --0.16$\pm$0.07  &   2 & --0.44$\pm$0.06  &   1 & --0.11$\pm$0.10  &  34 & --0.18$\pm$0.04  &   1 & --0.56$\pm$0.10 \\
1--40        &   4 & --0.05$\pm$0.10  &   2 & --0.48$\pm$0.06  &   2 &   0.07$\pm$0.15  &  25 & --0.16$\pm$0.05  &   1 & --0.57$\pm$0.10 \\
G18484\_0316 &   4 & --0.02$\pm$0.05  &   2 & --0.32$\pm$0.05  &   1 &   0.05$\pm$0.10  &  25 & --0.08$\pm$0.05  &   1 & --0.63$\pm$0.10 \\
1--110       &   0 &   \nodata        &   0 &   \nodata        &   1 & --0.17$\pm$0.10  &  16 & --0.17$\pm$0.07  &   0 &   \nodata       \\
G18484\_0309 &   1 & --0.19$\pm$0.10  &   0 &   \nodata        &   1 & --0.26$\pm$0.10  &  17 & --0.17$\pm$0.07  &   0 &   \nodata       \\
G18450\_0453 &   1 & --0.02$\pm$0.10  &   2 & --0.38$\pm$0.10  &   0 &   \nodata        &  15 & --0.13$\pm$0.07  &   0 &   \nodata       \\
1--32        &   1 & --0.35$\pm$0.10  &   1 & --0.41$\pm$0.10  &   0 &   \nodata        &  13 & --0.07$\pm$0.06  &   0 &   \nodata       \\
G18458\_0547 &   1 & --0.41$\pm$0.10  &   0 &   \nodata        &   0 &   \nodata        &  12 & --0.02$\pm$0.10  &   0 &   \nodata       \\
G18447\_0453 &   1 & --0.50$\pm$0.10  &   0 &   \nodata        &   0 &   \nodata        &   6 & --0.02$\pm$0.19  &   0 &   \nodata       \\
G18144\_0228 &   0 &   \nodata        &   0 &   \nodata        &   0 &   \nodata        &   5 & --0.22$\pm$0.15  &   0 &   \nodata       \\
G18564\_0457 &   0 &   \nodata        &   0 &   \nodata        &   0 &   \nodata        &   7 & --0.27$\pm$0.15  &   0 &   \nodata       \\
G18445\_0448 &   1 & --0.03$\pm$0.10  &   0 &   \nodata        &   0 &   \nodata        &   9 & --0.04$\pm$0.13  &   0 &   \nodata       \\
G18483\_0608 &   0 &   \nodata        &   0 &   \nodata        &   0 &   \nodata        &   6 & --0.25$\pm$0.15  &   0 &   \nodata       \\
G18279\_0101 &   0 &   \nodata        &   0 &   \nodata        &   0 &   \nodata        &   4 & --0.43$\pm$0.27  &   0 &   \nodata       \\
G18579\_0455 &   0 &   \nodata        &   0 &   \nodata        &   0 &   \nodata        &   0 &   \nodata        &   0 &   \nodata       \\
G18152\_0232 &   0 &   \nodata        &   0 &   \nodata        &   0 &   \nodata        &   0 &   \nodata        &   0 &   \nodata       \\
G18487\_0606 &   0 &   \nodata        &   0 &   \nodata        &   0 &   \nodata        &   0 &   \nodata        &   0 &   \nodata       \\
G18172\_0750 &   0 &   \nodata        &   0 &   \nodata        &   0 &   \nodata        &   0 &   \nodata        &   0 &   \nodata       \\
G18279\_0107 &   0 &   \nodata        &   0 &   \nodata        &   0 &   \nodata        &   0 &   \nodata        &   0 &   \nodata       \\
G18175\_0749 &   0 &   \nodata        &   0 &   \nodata        &   0 &   \nodata        &   0 &   \nodata        &   0 &   \nodata       \\
$<$MS$>$     &   0 &   \nodata        &   0 &   \nodata        &   0 &   \nodata        &   0 &   \nodata        &   0 &   \nodata       \\
\enddata                                          
\end{deluxetable}                                 
\clearpage                      
 
\begin{deluxetable}{lcccccccccc}
\rotate                         
\small                         
\footnotesize                  
\scriptsize                    
\tablenum{5d}                    
\tablewidth{0pt}                
\tablecaption{Abundance Ratios: Zn -- Eu
\label{tab_ratio_d}} 
\tablehead{                     
\colhead{Star} &                
\colhead{${\rm N_{Zn}}$} & \colhead{[Zn/Fe]} &
\colhead{${\rm N_{Zr}}$} & \colhead{[Zr/Fe]} &
\colhead{${\rm N_{Ba}}$} & \colhead{[Ba/Fe]} &
\colhead{${\rm N_{La}}$} & \colhead{[La/Fe]} &
\colhead{${\rm N_{Eu}}$} & \colhead{[Eu/Fe]} }
\startdata
IV--81       &   0 &   \nodata        &   3 &   0.19$\pm$0.05  &   3 &   0.03$\pm$0.05  &   2 &   0.23$\pm$0.06  &   1 &   0.61$\pm$0.10 \\
IV--59       &   1 &   0.43$\pm$0.10  &   2 & --0.23$\pm$0.05  &   3 &   0.22$\pm$0.09  &   0 &   \nodata        &   1 &   0.54$\pm$0.10 \\
IV--82       &   0 &   \nodata        &   2 & --0.25$\pm$0.05  &   3 & --0.01$\pm$0.05  &   1 &   0.22$\pm$0.10  &   1 &   0.55$\pm$0.10 \\
1--109       &   0 &   \nodata        &   0 &   \nodata        &   3 & --0.13$\pm$0.07  &   0 &   \nodata        &   0 &   \nodata       \\
1--36        &   0 &   \nodata        &   0 &   \nodata        &   3 &   0.01$\pm$0.05  &   1 & --0.01$\pm$0.10  &   1 &   0.60$\pm$0.10 \\
1--31        &   0 &   \nodata        &   0 &   \nodata        &   3 & --0.11$\pm$0.05  &   0 &   \nodata        &   1 &   0.53$\pm$0.10 \\
1--40        &   0 &   \nodata        &   0 &   \nodata        &   3 & --0.03$\pm$0.05  &   0 &   \nodata        &   0 &   \nodata       \\
G18484\_0316 &   0 &   \nodata        &   0 &   \nodata        &   3 & --0.14$\pm$0.05  &   0 &   \nodata        &   0 &   \nodata       \\
1--110       &   0 &   \nodata        &   0 &   \nodata        &   3 & --0.09$\pm$0.07  &   0 &   \nodata        &   0 &   \nodata       \\
G18484\_0309 &   0 &   \nodata        &   0 &   \nodata        &   3 & --0.11$\pm$0.05  &   0 &   \nodata        &   1 &   0.52$\pm$0.10 \\
G18450\_0453 &   0 &   \nodata        &   0 &   \nodata        &   3 &   0.04$\pm$0.05  &   0 &   \nodata        &   0 &   \nodata       \\
1--32        &   0 &   \nodata        &   0 &   \nodata        &   3 &   0.02$\pm$0.05  &   0 &   \nodata        &   0 &   \nodata       \\
G18458\_0547 &   0 &   \nodata        &   0 &   \nodata        &   3 &   0.04$\pm$0.05  &   0 &   \nodata        &   0 &   \nodata       \\
G18447\_0453 &   0 &   \nodata        &   0 &   \nodata        &   3 & --0.09$\pm$0.09  &   0 &   \nodata        &   0 &   \nodata       \\
G18144\_0228 &   0 &   \nodata        &   0 &   \nodata        &   3 & --0.12$\pm$0.13  &   0 &   \nodata        &   0 &   \nodata       \\
G18564\_0457 &   0 &   \nodata        &   0 &   \nodata        &   3 & --0.27$\pm$0.05  &   0 &   \nodata        &   0 &   \nodata       \\
G18445\_0448 &   0 &   \nodata        &   0 &   \nodata        &   3 & --0.05$\pm$0.05  &   0 &   \nodata        &   0 &   \nodata       \\
G18483\_0608 &   0 &   \nodata        &   0 &   \nodata        &   3 & --0.17$\pm$0.05  &   0 &   \nodata        &   0 &   \nodata       \\
G18279\_0101 &   0 &   \nodata        &   0 &   \nodata        &   3 & --0.22$\pm$0.10  &   0 &   \nodata        &   0 &   \nodata       \\
G18579\_0455 &   0 &   \nodata        &   0 &   \nodata        &   2 & --0.31$\pm$0.10  &   0 &   \nodata        &   0 &   \nodata       \\
G18152\_0232 &   0 &   \nodata        &   0 &   \nodata        &   3 & --0.47$\pm$0.17  &   0 &   \nodata        &   0 &   \nodata       \\
G18487\_0606 &   0 &   \nodata        &   0 &   \nodata        &   2 & --0.22$\pm$0.17  &   0 &   \nodata        &   0 &   \nodata       \\
G18172\_0750 &   0 &   \nodata        &   0 &   \nodata        &   1 & --0.65$\pm$0.11  &   0 &   \nodata        &   0 &   \nodata       \\
G18279\_0107 &   0 &   \nodata        &   0 &   \nodata        &   2 &   0.02$\pm$0.15  &   0 &   \nodata        &   0 &   \nodata       \\
G18175\_0749 &   0 &   \nodata        &   0 &   \nodata        &   0 &   \nodata        &   0 &   \nodata        &   0 &   \nodata       \\
$<$MS$>$     &   0 &   \nodata        &   0 &   \nodata        &   0 &   \nodata        &   0 &   \nodata        &   0 &   \nodata       \\
\enddata                                          
\end{deluxetable}                                 

%
%Tabla6
%
\begin{deluxetable}{lrrrrr}                       
%\rotate                                          
%\small                                           
%\footnotesize                                    
%\scriptsize                                      
\tablenum{6}                                      
\tablewidth{0pt}                                  
\tablecaption{Sensitivity of Abundance
\label{tab_sensit}}           
\tablehead{                                       
\colhead{} &                                      
\colhead{$\Delta$\ew} &                          
\colhead{$\Delta$\teff} &                        
\colhead{$\Delta$\grav} &                        
\colhead{$\Delta$\mtv} &                         
\colhead{$\Delta$\fe}\\                        
\colhead{} &                                      
\colhead{10\%} &                                  
\colhead{+ 100 K} &                               
\colhead{+ 0.2 dex} &                             
\colhead{+ 0.3 \kms} &                            
\colhead{+ 0.2 dex} }                             
\startdata                                        
Fe I :  & & & & &  \\
4250/1.5/1.8 &  0.10 &  0.06 &  0.02 & $-$0.09 & $-$0.01 \\
5000/3.0/1.5 &  0.10 &  0.09 &  0.00 & $-$0.07 &  0.00 \\
6000/4.0/1.0 &  0.11 &  0.08 & $-$0.02 & $-$0.06 &  0.00 \\
Fe II:  & & & & &  \\
4250/1.5/1.8 &  0.06 & $-$0.11 &  0.10 & $-$0.03 & $-$0.06 \\
5000/3.0/1.5 &  0.06 & $-$0.04 &  0.09 & $-$0.03 & $-$0.05 \\
O I(permitted): & & & & &  \\
4250/1.5/1.8 &  0.08 & $-$0.20 &  0.10 & $-$0.02 & $-$0.01 \\
5000/3.0/1.5 &  0.07 & $-$0.12 &  0.07 & $-$0.02 & $-$0.01 \\
O I(forbidden): & & & & &  \\
4250/1.5/1.8 &  0.05 &  0.02 &  0.08 &  0.00 & $-$0.07 \\
5000/3.0/1.5 &  0.05 &  0.03 &  0.07 &  0.00 & $-$0.06 \\
Na I :  & & & & &  \\
4250/1.5/1.8 &  0.14 &  0.14 & $-$0.03 & $-$0.05 &  0.00 \\
5000/3.0/1.5 &  0.09 &  0.11 & $-$0.05 & $-$0.02 & $-$0.01 \\
6000/4.0/1.0 &  0.09 &  0.08 & $-$0.05 &  0.00 &  0.00 \\
Mg I :  & & & & &  \\
4250/1.5/1.8 &  0.19 &  0.06 &  0.00 & $-$0.13 &  0.01 \\
5000/3.0/1.5 &  0.12 &  0.08 & $-$0.02 & $-$0.06 &  0.00 \\
Si I :  & & & & &  \\
4250/1.5/1.8 &  0.07 & $-$0.04 &  0.04 & $-$0.03 & $-$0.03 \\
5000/3.0/1.5 &  0.06 &  0.01 &  0.01 & $-$0.01 & $-$0.01 \\
Ca I :  & & & & &  \\
4250/1.5/1.8 &  0.16 &  0.13 & $-$0.02 & $-$0.13 &  0.02 \\
5000/3.0/1.5 &  0.11 &  0.09 & $-$0.03 & $-$0.06 &  0.01 \\
6000/4.0/1.0 &  0.11 &  0.08 & $-$0.06 & $-$0.06 &  0.00 \\
Sc II:  & & & & &  \\
4250/1.5/1.8 &  0.11 & $-$0.02 &  0.08 & $-$0.07 & $-$0.06 \\
5000/3.0/1.5 &  0.07 &  0.00 &  0.08 & $-$0.03 & $-$0.05 \\
Ti I :  & & & & &  \\
4250/1.5/1.8 &  0.07 &  0.17 &  0.00 & $-$0.03 &  0.01 \\
5000/3.0/1.5 &  0.06 &  0.12 &  0.00 & $-$0.01 &  0.01 \\
V I  :  & & & & &  \\
4250/1.5/1.8 &  0.07 &  0.20 &  0.00 & $-$0.03 &  0.01 \\
5000/3.0/1.5 &  0.04 &  0.15 &  0.00 &  0.00 &  0.01 \\
Cr I :  & & & & &  \\
4250/1.5/1.8 &  0.08 &  0.11 &  0.00 & $-$0.01 &  0.01 \\
5000/3.0/1.5 &  0.16 &  0.15 & $-$0.02 & $-$0.13 &  0.02 \\
Mn I :  & & & & &  \\
4250/1.5/1.8 &  0.08 &  0.14 &  0.00 & $-$0.01 &  0.00 \\
5000/3.0/1.5 &  0.05 &  0.09 &  0.00 &  0.00 &  0.00 \\
Co I :  & & & & &  \\
4250/1.5/1.8 &  0.06 &  0.08 &  0.03 & $-$0.01 & $-$0.02 \\
5000/3.0/1.5 &  0.06 &  0.11 &  0.01 &  0.00 &  0.00 \\
Ni I :  & & & & &  \\
4250/1.5/1.8 &  0.09 &  0.03 &  0.03 & $-$0.06 & $-$0.03 \\
5000/3.0/1.5 &  0.07 &  0.08 &  0.01 & $-$0.03 &  0.00 \\
Cu I :  & & & & &  \\
4250/1.5/1.8 &  0.07 &  0.08 &  0.03 & $-$0.01 & $-$0.02 \\
5000/3.0/1.5 &  0.05 &  0.10 &  0.01 &  0.00 &  0.00 \\
Zn I :  & & & & &  \\
4250/1.5/1.8 &  0.05 & $-$0.08 &  0.06 & $-$0.01 & $-$0.03 \\
Zr I :  & & & & &  \\
4250/1.5/1.8 &  0.04 &  0.22 &  0.00 & $-$0.01 &  0.01 \\
Ba II:  & & & & &  \\
4250/1.5/1.8 &  0.23 &  0.02 &  0.07 & $-$0.25 & $-$0.06 \\
5000/3.0/1.5 &  0.15 &  0.02 &  0.07 & $-$0.16 & $-$0.06 \\
6000/4.0/1.0 &  0.12 &  0.05 &  0.05 & $-$0.13 & $-$0.01 \\
La II:  & & & & &  \\
4250/1.5/1.8 &  0.06 &  0.01 &  0.08 & $-$0.02 & $-$0.07 \\
Eu II:  & & & & &  \\
4250/1.5/1.8 &  0.06 & $-$0.02 &  0.09 & $-$0.02 & $-$0.07 \\
5000/3.0/1.5 &  0.05 &  0.00 &  0.08 & $-$0.02 & $-$0.06 
\enddata                                          
\end{deluxetable}

%
% Tabla7
%
\clearpage
\begin{deluxetable}{lrlcc}
\tablenum{7}
\tablewidth{0pt}
%\small
%\footnotesize
%\scriptsize
\tablecaption{Mean Iron Abundance and Abundance Ratios. 
\label{tab_mean}}
\tablehead{\colhead{} & \colhead{\# stars} & \colhead{$<$[X/Fe]$>$} 
& \colhead{$\sigma_{obs}$} & \colhead{$\sigma_{pred}$} \\
\colhead{} & \colhead{} & \colhead{(dex)} & \colhead{(dex)} &
\colhead{(dex)} }
\startdata
Fe I \tablenotemark{a} & 25 &--1.30$\pm$0.02 &  0.12 &  0.14  \\
Fe II\tablenotemark{a} & 17 &--1.28$\pm$0.02 &  0.08 &  0.14  \\
O I                    &  9 & +0.26$\pm$0.03 &  0.10 &  0.05  \\
Na I                   & 25 & +0.08$\pm$0.05 &  0.26 &  0.12  \\
Mg I                   & 13 & +0.29$\pm$0.02 &  0.09 &  0.12  \\
Si I                   & 17 & +0.21$\pm$0.02 &  0.07 &  0.07  \\
Ca I                   & 25 & +0.33$\pm$0.03 &  0.14 &  0.08  \\
Sc II                  & 19 & +0.05$\pm$0.03 &  0.13 &  0.14  \\
Ti I                   & 19 & +0.16$\pm$0.02 &  0.10 &  0.11  \\
V I                    & 11 &--0.14$\pm$0.04 &  0.12 &  0.13  \\
Cr I                   & 14 &--0.13$\pm$0.05 &  0.17 &  0.15  \\
Mn I                   &  9 &--0.39$\pm$0.03 &  0.10 &  0.11  \\
Co I                   & 10 &--0.06$\pm$0.03 &  0.11 &  0.09  \\
Ni I                   & 19 &--0.12$\pm$0.02 &  0.10 &  0.05  \\
Cu I                   &  7 &--0.53$\pm$0.02 &  0.05 &  0.10  \\
Zn I                   &  1 & +0.43          &  ...  &  0.08  \\
Zr I                   &  3 &--0.10$\pm$0.12 &  0.20 &  0.19  \\
Ba II                  & 24 &--0.08$\pm$0.04 &  0.13 &  0.18  \\
La II                  &  3 & +0.18$\pm$0.07 &  0.11 &  0.14  \\
Eu II                  &  6 & +0.56$\pm$0.02 &  0.04 &  0.16  \\
\enddata
\tablenotetext{a}{For Fe, [Fe/H] is given.  For
all other elements, [X/Fe] is given.}
\end{deluxetable}

%
% Tabla8
%
\clearpage
\begin{deluxetable}{lcccc}
\tablenum{8}
\tablewidth{0pt}
%\small
%\footnotesize
%\scriptsize
\tablecaption{Comparison of Our Abundance Analysis For M5 With Earlier Work
\label{tab_comp}}
\tablehead{\colhead{[X/Fe]} 
& \colhead{\citet{sne92}}
& \colhead{\citet{iva01}} 
& \colhead{This work}
& \colhead{$\sigma_c$} \\
\colhead{} & \colhead{(dex)} & \colhead{(dex)} & \colhead{(dex)} &
\colhead{(dex)} 
}
\startdata
Fe I  \tablenotemark{a} 
      & --1.17$\pm$0.01 & --1.34$\pm$0.01 & --1.30$\pm$0.02 & 0.11 \\
Fe II \tablenotemark{a} 
      & --1.16$\pm$0.02 & --1.21$\pm$0.01 & --1.28$\pm$0.02 & 0.14 \\
O I   &  +0.11$\pm$0.06 &  +0.02$\pm$0.04 &  +0.26$\pm$0.03 & 0.03 \\
Na I  & --0.07$\pm$0.06 &  +0.11$\pm$0.03 &  +0.08$\pm$0.05 & 0.08 \\
Mg I  &  \nodata        &  +0.34$\pm$0.03 &  +0.29$\pm$0.02 & 0.04 \\
Si I  &  +0.20$\pm$0.02 &  +0.31$\pm$0.01 &  +0.21$\pm$0.02 & 0.06 \\
Ca I  &  +0.19$\pm$0.01 &  +0.26$\pm$0.01 &  +0.33$\pm$0.03 & 0.05 \\
Sc II & --0.10$\pm$0.03 & --0.01$\pm$0.02 &  +0.05$\pm$0.03 & 0.13 \\
Ti I  &  +0.29$\pm$0.04 &  +0.22$\pm$0.02 &  +0.16$\pm$0.02 & 0.10 \\
V  I  &  +0.05$\pm$0.03 & --0.10$\pm$0.02 & --0.14$\pm$0.04 & 0.12 \\
Mn I  &  \nodata        & --0.25$\pm$0.02 & --0.39$\pm$0.03 & 0.09 \\
Ni I  & --0.10$\pm$0.02 & --0.05$\pm$0.01 & --0.12$\pm$0.02 & 0.05 \\
Ba II &  \nodata        &  +0.16$\pm$0.02 & --0.08$\pm$0.04 & 0.15 \\
La II &  \nodata        &  +0.02$\pm$0.02 &  +0.18$\pm$0.07 & 0.12 \\
Eu II &  \nodata        &  +0.43$\pm$0.02 &  +0.56$\pm$0.02 & 0.15
\enddata
\tablenotetext{a}{For Fe, [Fe/H] is given.  For
all other elements, [X/Fe] is given.}
\end{deluxetable}

%
% Tabla9
%
\clearpage
\begin{deluxetable}{lcccc}
\tablenum{9}
\tablewidth{0pt}
%\small
%\footnotesize
%\scriptsize
\tablecaption{Updated Mean Iron Abundance and Abundance Ratios in M71.
\label{table_newm71}}
\tablehead{\colhead{} & \colhead{\# stars} & \colhead{$<$[X/Fe]$>$}
& \colhead{$\sigma_{obs}$} & \colhead{$\sigma_{pred}$} \\
\colhead{} & \colhead{} & \colhead{(dex)} & \colhead{(dex)} &
\colhead{(dex)} }
\startdata
Fe I \tablenotemark{a} & 25 &--0.63$\pm$0.02 &  0.08 &  0.16  \\
Fe II\tablenotemark{a} & 25 &--0.73$\pm$0.03 &  0.13 &  0.21  \\
C I                    &  6 & +1.36$\pm$0.20 &  0.49 &  0.13  \\
O I                    & 25 & +0.10$\pm$0.04 &  0.18 &  0.12  \\
Na I                   & 25 & +0.10$\pm$0.03 &  0.14 &  0.10  \\
Mg I                   & 24 & +0.35$\pm$0.02 &  0.10 &  0.12  \\
Al I                   & 11 & +0.22$\pm$0.03 &  0.10 &  0.10  \\
Si I                   & 24 & +0.27$\pm$0.03 &  0.14 &  0.17  \\
K I                    & 11 &--0.12$\pm$0.09 &  0.30 &  0.29  \\
Ca I                   & 25 & +0.36$\pm$0.02 &  0.09 &  0.11  \\
Sc II                  & 25 & +0.03$\pm$0.03 &  0.16 &  0.15  \\
Ti I                   & 25 & +0.23$\pm$0.02 &  0.12 &  0.11  \\
V I                    & 21 & +0.04$\pm$0.02 &  0.14 &  0.17  \\
Cr I                   & 23 &--0.07$\pm$0.02 &  0.09 &  0.09  \\
Mn I                   & 13 &--0.28$\pm$0.03 &  0.11 &  0.11  \\
Co I                   & 17 &--0.07$\pm$0.01 &  0.05 &  0.10  \\
Ni I                   & 25 & +0.01$\pm$0.01 &  0.06 &  0.09  \\
Cu I                   & 21 &--0.07$\pm$0.03 &  0.14 &  0.21  \\
Zn I                   &  8 & +0.40$\pm$0.06 &  0.16 &  0.22  \\
Y II                   &  3 &--0.07$\pm$0.02 &  0.04 &  0.15  \\
Zr I                   & 10 &--0.04$\pm$0.08 &  0.25 &  0.16  \\
Ba II                  & 25 & +0.19$\pm$0.02 &  0.12 &  0.19  \\
La II                  & 14 & +0.13$\pm$0.03 &  0.10 &  0.15  \\
Eu II                  & 11 & +0.36$\pm$0.05 &  0.31 &  0.11  \\
\enddata
\tablenotetext{a}{For Fe, [Fe/H] is given.  For
all other elements, [X/Fe] is given.}
\end{deluxetable}


\begin{thebibliography}{}

%\bibitem[Allende Prieto, Lambert \& Asplund(2001)]{all01} Allende Prieto, C.,
%Lambert, D. L., \& Asplund, M., 2001, \apjl, 556, L63

\bibitem[Anders \& Grevesse(1989)]{and89} Anders, E. \& Grevesse, N., 1989,
Geochim. Cosmochim. Acta, 53, 197

\bibitem[Arp(1962)]{arp62} Arp, H. C., 1962, \aj, 135, 311

\bibitem[Arneson \etal(1977)]{arneson77} 
Aresen, A., Bengtsson., A., Hallin, R., Lindskog, J. \& Nordland, T.,
1977, Physica Scripta, 16, 31

\bibitem[Auman \& Woodrow(1975)]{aum75} Auman, J. R. \& Woodrow, J. E. J.,
1975, \apj, 197, 163

%\bibitem[Arp \& Hartwick(1971)]{arp71} Arp, H. C. \& Hartwick, F. D. A., 1971,
%\apj, 167, 499

\bibitem[Bard \etal(1991)]{bar91} Bard, A., Kock, A., \& Kock, M.,
1991, \aap, 248, 315

\bibitem[Bard \& Kock(1994)]{bar94} Bard, A., \& Kock, M., 1994,
\aap, 282, 1014

\bibitem[Basu, Pinssoneault \& Bahcall(2000)]{bas00} Basu, S., Pinsonneault,
M.~H. \& Bahcall, J.~N., 2000, \apj, 529, 1081

%\bibitem[Baum\"uller \& Gehren(1996)]{bau96} Baum\"uller, D. \&
%Gehren, T., 1996, \aap, 307, 961
%
%\bibitem[Baum\"uller \& Gehren(1997)]{bau97} Baum\"uller, D. \& Gehren, T.,
%1997, \aap, 325, 1088

\bibitem[Baum\"uller \etal(1998)]{bau98} Baum\"uller, D., Butler, K., \&
Gehren, T., 1998, \aap, 338, 637

\bibitem[Behr \etal(1999)]{beh99} 
Behr, B.~B., Cohen, J.~G. \& McCarthy, J.~K., 1999, \apjl, 517, L135

\bibitem[Behr, Cohen \& McCarthy(2000)]{beh00} 
Behr, B.~B., Cohen, J.~G. \& McCarthy, J.~K., 2000, \apjl, 531, L37

\bibitem[Bi\'emont \etal(1991a)]{bie91a} Bi\'emont, E., Baudoux, M.,
Kurucz, R. L., Ansbacher, W., \& Pinnimgton, E. H., 1991a, \aap, 249, 539

%\bibitem[Bi\'emont \etal(1991)]{bie91} Bi\'emont, E., Hibbert, A., 
%Godefroid, M., Vaeck, N., \& Fawcett, B. C., 1991, \apj, 375, 818

\bibitem[Bi\'emont \etal(1981)]{biemont81} 
Bi\'emont, E., Grevesse, N., Hannaford, P. \& Lowe, R.M.,  
1981, \apj, 248, 867

\bibitem[Blackwell \etal(1979)]{bla79} Blackwell, D. E., Petford, A. D.,
\& Shallis, M. J., 1979, \mnras, 186, 657

\bibitem[Blackwell \etal(1980)]{bla80} Blackwell, D. E., Shallis, M. J.,
\& Simmons, G. J., 1980, \aap, 81, 340

\bibitem[Blackwell \etal(1982a)]{bla82a} Blackwell, D. E., Petford, A. D.,
Shallis, M. J., \& Simmons, G. J., 1982$a$, \mnras, 199, 43

\bibitem[Blackwell \etal(1982b)]{bla82b} Blackwell, D. E., Petford, A. D.,
\& Simmons, G. J., 1982$b$, \mnras, 201, 595

\bibitem[Blackwell \etal(1986)]{bla86} Blackwell, D. E., Booth, A. J.,
Haddock, D. J., Petford, A. D., \& Leggett, S. K., 1986, \mnras, 220, 549

\bibitem[Bonifacio \etal(2002)]{bon02} Bonifacio, P. \etal, 2002,
\aap, 390, 91 

\bibitem[Bono \etal(2001)]{bon01} Bono, G., Cassisi, S., Zocalli, M. \& 
Piotto, G., 2001, \apj, 546, L109

\bibitem[Briley \& Cohen(2001)]{bri01a} Briley, M. M. \& Cohen, J. G., 2001,

\aj, 122, 242
%
%\bibitem[Briley \etal(2001)]{bri01b} Briley, M. M., Smith, G. H., \& 
%Claver, C. F., 2001, \aj, 122, 2561
%
%\bibitem[Brown \etal(1990)]{bro90} Brown J. A., Wallerstein, G., \&
%Oke, J. B., 1990, \aj, 100, 1561
%
%\bibitem[Brown \& Wallerstein(1992)]{bro92} Brown J. A. \& Wallerstein, G.,
%\aj, 104, 1818

\bibitem[Buonanno \etal(1981)]{buo81} Buonanno, R., Corsi, C.~E. \& 
Fusi Pecci, F., 1981, \mnras, 196, 435                  

\bibitem[Burris \etal(2000)]{bur00} Burris, D. L., Pilachowski, C. A.,
Armandroff, T. E., Sneden, C., Cowan, J. J., \& Roe, H., 2000, \apj, 544, 302

%\bibitem[Cameron(1982)]{cam82} Cameron, A. G. W., 1982, Ap.Sp.Sci., 82, 123

\bibitem[Cannon \etal(1998)]{can98} Cannon, R. D., Croke, B. F. W., Bell, R. A.,
 Hesser, J. E., \& Stathakis, R. A., 1998, \mnras, 298, 601

%\bibitem[Castilho \etal(2000)]{cas00} Castilho, B. V., Pasquini, L., Allen, D. M., 
%Barbuy, B., \& Molaro, P., 2000, \aap, 361, 92

\bibitem[Castellani \etal(1997)]{cas97} Castellani, V.,
Ciacio, F., Delg'Innocenti, S. \& Fiorentini, G. 1997, \aap, 322, 801

\bibitem[Cavallo \etal(1998)]{cav98} Cavallo, R. M., Sweigart, A. V., \& 
Bell, R. A., 1998, \apj, 492, 575

\bibitem[Chaboyer \etal(2002)]{cha01} Chaboyer, B., Fenton, W.~H.,
Nelan, J.~E., Patnaude, D.~J. \& Simon, F.~E., 2002, \apj, 562 521

\bibitem[Charbonnel(1994)]{cha94} Charbonnel, C., 1994, \aap, 282, 811

\bibitem[Charbonnel(1995)]{cha95} Charbonnel, C., 1995, \apjl, 453, L4

%\bibitem[Charbonnel \& Balachandran(2000)]{cha00} Charbonnel, C. \& 
%Balachandran, S. C., 2000, \aap, 359, 563

\bibitem[Cleveland(1993)]{cle93} Cleveland, W.S., 1993, 
{\it Visualizing Data}, Hobart Press, Summit, New Jersey

%
%\bibitem[Cohen(1983)]{coh83} Cohen, J. G., 1983, \apj, 270, 654
%

\bibitem[Cohen(1999)]{coh99} Cohen, J. G., 1999, \aj, 117, 2434

\bibitem[Cohen \etal(1981)]{coh81} Cohen, J.~G., Frogel, J.~A., 
Persson, S.~E. \& Elias, J.~H., \apj, 249, 481, 1981

\bibitem[Cohen \etal(2001)]{coh01} Cohen, J. G., Behr, B. B., \& Briley, M. M., 
2001, \aj, 122, 1420 

\bibitem[Cohen, Briley \& Stetson(2002)]{coh02} Cohen, J. G., Briley, M. M.,
\& Stetson, P. B., 2002, \aj, 123, 2525

\bibitem[Corliss \& Bozman(1962)]{corliss62}
Coliss, C.~H. \& Bozman, W.~R., 1962, {\it{Experimental Transition
Probabilities for Spectral Lines of Seventy Elements}},
NBS Monograph 53, US Government Printing Office

\bibitem[Cudworth(1979)]{cud79} Cudworth, K.C., 1979, \aj, 84, 1866

%\bibitem[de la Reza \& Muller(1975)]{del75} de la Reza, R. \& Muller, E.A., 
%1975, olar Physics, 43, 15

\bibitem[Denisenkov \& Denisenkova(1990)]{den90} Denisenkov, P. A. \&
Denisenkova, S. N., SvAL, 16, 275 

\bibitem[Denissenkov \& Weiss(1996)]{den96} Denissenkov, P. A. \& Weiss, A.,
1996, \aap, 308, 773

%\bibitem[Denissenkov \& Weiss(2001)]{den01} Denissenkov, P. A. \& Weiss, A.,
%2001, \apj, 559, 115L

\bibitem[Fuhr \etal(1988)]{fuh88} Fuhr, J. R., Martin, G. A., \& Wiese, W. L.,
1988, J. Phys. Chem. Ref. Data 17, Suppl. 4 

%\bibitem[Fulbright(2000)]{ful00} Fulbright, J., 2000, \apj, 120, 1841
%

\bibitem[Gehren \etal(2001)]{geh01a} Gehren, T., Butler, K.,
Mashonkina, L., Reetz, J. \& Shi, J., 2001a, \aap, 366, 981

\bibitem[Gehren, Korn \& Shi(2001)]{geh01b} Gehren, T., Korn, A.~J.
\& Shi, J., 2001b, \aap, 380, 645

%\bibitem[Gratton \etal(1986)]{gra86} Gratton, R. G., Quarta, M. L.,
%\& Ortolani, S., 1986, \aap, 169, 208

\bibitem[Gratton \& Sneden(1994)]{gra94} Gratton, R. G. \& Sneden, C., 
1994, \aap, 287, 927

\bibitem[Gratton \etal(1999)]{gra99} Gratton, R. G., Carretta, E., Eriksson, K.,
\& Gustafsson, B., 1999, \aap, 350, 955

%\bibitem[Gratton \etal(2000)]{gra00} Gratton, R.G., Sneden, C., Carretta, E. 
%\& Bragaglia, A., 2000, \aap, 354, 169

\bibitem[Gratton \etal(2001)]{gra01} Gratton, R. G., Bonifacio, P.,
Bragaglia, A., Carretta, E., Castellani, V., Centurion, M., Chieffi, A.,
Claudi, R., Clementini, G., D'Antona, F., Desidera, S., Francois, P.,
Grundahl, F., Lucatello, S., Molaro, P., Pasquini, L., Sneden, C., Spite, F.,
\& Straniero, O., 2001, \aap, 369, 87

\bibitem[Grevesse \& Sauval(1998)]{gre98} Grevesse, N. \& Sauval, A. J., 
1998, Space Science Reviews, 85, 161

% \bibitem[Guenther, Kim \& Demarque(1996)]{gue96} Guenther, D.~B.,
% Kim, Y.~C. \& Demarque, P., 1996, \apj, 163, 382
% diffusion and solar helioseismology, replaced by Basu et al 2000

\bibitem[Gustafsson \etal(1975)]{gus75} Gustafsson, B., Bell, 
R.A., Eriksson, K. \& Nordlund, \aap, 1975, A\&A, 42, 407

\bibitem[Harris(1996)]{har96} Harris, W. E., 1996, \aj, 112, 1487

%\bibitem[Hinkle \etal(2000)]{hinkle00}
%Hinkle, K., Wallace, L., Valenti, J. \& Harmer, D., 2000,
%{\it{Visible and Near Infrared Atlas of the Arcturus Spectrum,
%3727 -- 9000 \AA}}, ASP Press, San Francisco
%%

\bibitem[Holweger \etal(1991)]{hol91} Holweger, H., Bard, A., Kock, A., \& 
Kock, M., 1991, \aap, 249, 545

\bibitem[Houdashelt, Bell \& Sweigart(2000)]{hou00}
Houdashelt, M.~L., Bell, R.~A. \& Sweigart, A.~V., 2000, \aj, 119, 1448

%\bibitem[Iben \& Renzini(1983)]{ibe83} Iben, I. Jr. \& Renzini, A., 1983, \araa,
%21, 271
%
%\bibitem[Israelian \etal(2001)]{isr01} Israelian, G., Rebolo, R., 
%Garc\'{\i}a L\'{o}pez, R. J., Bonifacio, P., Molaro, P., Basri, G., \&
%Shchukina, N., \apj, 551, 833

\bibitem[Ivans \etal(1999)]{iva99} Ivans, I. I., Sneden, C., Kraft, R. P., 
Suntzeff, N. B., Smith, V. V., Langer, G. E., \& Fulbright, J. P., 
1999, \apj, 118, 1273

\bibitem[Ivans \etal(2001)]{iva01} Ivans, I. I., Kraft, R. P., Sneden, C., 
Smith, G., Rich, R. M., \& Shetrone, M., 2001, \apj, 122, 1438

\bibitem[Johnson \& Bolte(1998)]{joh98} Johnson, K.~A.. \& Bolte, M., 1998,
\aj, 115, 693

\bibitem[Johnson(2002)]{joh02} Johnson, J., 2002, \apjs, 139, 219

%\bibitem[K\"appeler \etal(1989)]{kap89} K\"appeler, F., Beer, H., \& Wisshak, K., 
%1989, Rep.Prog.Phys., 52, 945

\bibitem[King \etal(1998)]{kin98} King, J. R., Stephens, A., Boesgaard, A. M.,
\& Deliyannis, C. P., 1998, \apj, 115, 666

\bibitem[Korn \& Gehren(2002)]{kor02} Korn, A.~J. \& Gehren, T., 2002,
in IAU Symposium 210, {\it Modelling of Stellar Atmospheres},
Uppsala, June 2002.

\bibitem[Kraft \etal(1993)]{kra93} Kraft, R. P., Sneden, C., Langer, G. E., \&
Shetrone, M. D., 1993, \aj, 106, 1490

\bibitem[Kraft(1994)]{kra94} Kraft, R. P., 1994, \pasp, 106, 553

%\bibitem[Kraft \etal(1995)]{kra95} Kraft, R. P., Sneden, C., Langer, G. E.,
%Shetrone, M. D., \& Bolte, M., 1995, \aj, 109, 2586
%
%\bibitem[Kraft \etal(1999)]{kra99} Kraft, R. P., Peterson, R. C., Puragra, G., 
%Sneden, C., Fulbright, J. P., \& Langer, G. E., 1999, \apj, 518, L53

\bibitem[Kurucz(1993a)]{kur93a} Kurucz, R. L., 1993$a$, ATLAS9 Stellar 
Atmosphere Programs and 2 km/s Grid, (Kurucz CD-ROM No. 13)

\bibitem[Kurucz(1993b)]{kur93b} Kurucz, R. L., 1993$b$, SYNTHE Spectrum 
Synthesis Programs and Line Data (Kurucz CD-ROM No. 18)

%\bibitem[Lambert(2002)]{lam02} Lambert, D. L., 2002, Highlights in Astronomy,
%in press
%
%%\bibitem[Lambert \etal(1996)]{lam96} Lambert, D. L., Heath, J. E., Lemke, M., 
%%\& Drake, J., 1996, \apjs, 103, 183

\bibitem[Lambert \& Allende-Prieto(2002)]{lambert02}
Lambert, D.~L. \& Allende-Prieto, C., 2002, \mnras, 335, 325

\bibitem[Langer \etal(1993)]{lan93} Langer, G. E., Hoffman. R. \& Sneden, C., 
1993, \pasp, 105, 301

%\bibitem[Langer \& Hoffman(1995)]{lan95} Langer, G. E., \& Hoffman, R. D.,
%1995, \pasp, 107, 1177
%
%\bibitem[Langer \etal(1997)]{lan97} Langer, G. E., Hoffman, R. D., \&
%Zaidins, C. S., 1997, \pasp, 109, 244

\bibitem[Lawler, Bonvallet \& Sneden(2001)]                              
{lawler01a} Lawler, J.~E., Bonvallet, G. \& Sneden, C., 2001, \apj, 556, 452 

\bibitem[Lawler \etal(2001)]{lawler01b}
Lawler, J.~E., Wickliffe, M.~E., den Hartog, E.~A. \& Sneden, C., 2001, \apj, 563, 1075

%\bibitem[Leep \etal(1987)]{lee87} Leep, E. M., Oke, J. B., \&
%Wallerstein, G., 1987, \aj, 93, 338

\bibitem[Martin \etal(1988)]{mar88} Martin, G. A., Fuhr, J. R., \& Wiese, W. L.,
1988, J. Phys. Chem. Ref. Data 17, Suppl. 3 

\bibitem[Mashonkina, Gehren \& Bikmaev(1999)]{mas99}
Mashonnkina, L., Gehren, T. \& Bikmaev, I., 1999, \aap, 343, 519

\bibitem[May \etal(1974)]{may74} May, M., Richter, J., \& Wichelmann, J.,
1974, \aaps, 18, 405

\bibitem[McWilliam(1997)]{mcw97} McWilliam, A., 1997, \araa, 35, 503

\bibitem[McWilliam(1998)]{macwilliam98} McWilliam, A., 1998, \aj, 115, 1640

%\bibitem[Mel\'endez \etal(2001)]{mel01} Mel\'endez, J., Barbuy, B., 
%\& Spite, F., 2001, \apj, 556, 858

\bibitem[Michaud, Fontaine \& Beaudet(1981)]{mic81} Michaud, G., Fontaine,
G. \& Beaudet, G., 1984, \apj, 282, 206

\bibitem[Moore \etal(1966)]{moo66} Moore, C. E., Minnaert, M. G. J., \& 
Houtgast, J.,
1966,  {\it{The Solar Spectrum 2935 \AA\ to 8770 \AA}}, 
National Bureau of Standards Monograph, 
Washington: US Government Printing Office (USGPO).

%\bibitem[Norris \& DaCosta(1995)]{nor95} Norris \& DaCosta, 1995, \apj, 447, 680

\bibitem[O'Brian \etal(1991)]{obr91} O'Brian, T. R., Wickliffe, M. E.,
Lawler, J. E., Whaling, W., \& Brault, J. W., 1991, J. Opt. Soc. Am., B8, 1185

%\bibitem[Peterson(1980)]{pet80} Peterson, R. C., 1980, \apj, 237, 87

\bibitem[Pinsonneault(1997)]{pin97} Pinsonneault, M., 1997, \araa, 35, 557

\bibitem[Prochaska \etal(2000)]{pro00} Prochascka, J. X., Naumov, S. O., 
Carney, B. W., McWilliam, A., \& Wolfe, A. M., 2000, \apj, 120, 2513 

\bibitem[Ram\'{\i}rez \etal(2001)]{ram01} Ram\'{\i}rez, S. V., Cohen, J. G., 
Buss, J., \& Briley, M. M., 2001, \aj, 122, 1429

\bibitem[Ram\'{\i}rez \& Cohen(2002)]{ram02} Ram\'{\i}rez, S. V. \& 
Cohen, J. G., 2002, \aj, 123, 3277

%\bibitem[Ryan \etal(2001)]{rya01} Ryan, S. G., Kajino, T., Beers, T. C., 
%Suzuki, T. K., Romano, D., Matteucci, F. \& Rosolankova, K., 2001, \apj, 545, 55

\bibitem[Rastorguev \& Samus(1991)]{rastor91} Rastroguev, A.S. \& Samus, N.N.,
1991, Soviet Astron. Lett., 17, 388

\bibitem[Richard \etal(2002)]{ric02} Richard, P., Michaud, G.,
Richer, J., Turcotte, S., Turck-Chieze, S. \& VandenBerg, D.~A.,
2002, \apj, 568, 979
% Models of Metal poor stars with gravitational settling and
% radiative accelerations, 1. evolution and abundance anomalies

\bibitem[Rutten(1978)]{rut78} Rutten, R.~J., 1978, Solar Physics, 56, 237

\bibitem[Salaris, Groenwegen \& Weiss(2000)]{sal00} Salaris, M.,
Groenwegen, M.~A.~T. \& Weiss, A., 2000, \aap, 355, 299

\bibitem[Salaris \& Weiss(2001)]{sal01} Salaris, M. \& Weiss, A., 2001, 
\aap, 376, 955

\bibitem[Sandquist \etal(1996)]{san96} Sandquist, E.~L., Bolte, M., 
Stetson, P.~B., Hesser, J.~E., 1996, \apj, 470, 910

\bibitem[Schlegel, Finkbeiner \& Davis(1998)]{sch98} Schlegel, D.~J., 
Finkbeiner, D.~P. \& Davis, M., 1998, \apj, 500, 525

\bibitem[Shetrone(1996)]{she96} Shetrone, M. D., 1996, \aj 112, 1517

%\bibitem[Shetrone \etal(1993)]{she93} Shetrone, M. D., Sneden, C. \& 
%Pilachowski, C. A., 1993, \pasp, 105, 337

\bibitem[Smith(1981)]{smi81a} Smith, G., 1981, \aap, 103, 351

\bibitem[Smith \& Raggett(1981)]{smi81} Smith, G. \& Raggett, D. St. J., 1981,
J. Ph. B, 14, 4015

\bibitem[Sneden(1973)]{sne73} Sneden, C., 1973, Ph.D. thesis, Univ. of Texas

%%\bibitem[Sneden \etal(1991)]{sne91} Sneden, C., Kraft, R. P., Prosser, C. F.,
%%\& Langer, G. E., 1991, \aj, 102, 2001

\bibitem[Sneden \etal(1992)]{sne92} Sneden, C., Kraft, R. P., Langer, G. E.,
Prosser, C. F., \& Shetrone, M. D., 1992, \aj, 104, 2121

%\bibitem[Sneden \etal(1994)]{sne94} Sneden, C., Kraft, R. P., Langer, G. E.,
%Prosser, C. F., \& Shetrone, M. D., 1994, \aj, 107, 1773

\bibitem[Sneden \etal(1997)]{sne97} Sneden, C., Kraft, R. P., Langer, G. E.,
Prosser, C. F., \& Shetrone, M. D., 1997, \aj, 114, 1964

\bibitem[Sneden \etal(2001)]{sne01} Sneden, C., Cowan, J. J. \& Truran, J. W., 
2001, Astro-ph/0101439

%\bibitem[Spite \& Spite(1982)]{spi82} Spite, F. \& Spite, M., 1982, 
%\aap, 115, 357

\bibitem[Stetson \etal(1998)]{ste98} Stetson, P. B., Hesser, J. E., 
Smecker-Hane, T. A., 1998, \pasp, 110, 533

\bibitem[Stetson \etal(1999)]{ste99} Stetson, P.~B., Bolte, M., 
Harris, W.~E., Hesser, J.~E.,
van den Bergh, S., VandenBerg, D.~A., Bell, R.~A., Johnson, J.~A.,
Bond, H.~E., Fullton, L.~K., Fahlman, G.~C. \& Richer, H.~B.,
1999, \aj, 117, 247

\bibitem[Stetson(2000)]{ste00} Stetson, P. B., 2000, \pasp, 112, 925

%%\bibitem[Suntzeff \& smith(1991)]{sun91} Suntzeff, N. B. \& Smith, V. V., 
%%1991, \apj, 381, 160

\bibitem[Sweigart \& Mengel(1979)]{swe79} Sweigart, A. V. \& Mengel, J. G., 
1979, \apj, 229, 624

%\bibitem[Takeda \etal(2001)]{tak01} Takeda, Y., Zhao, G., Chen, Y., Qiu, H., \&
%Takada-Hidai, M., 2000, \pasj, submitted (astro-ph/0110165)
%
%\bibitem[Takeda \etal(2002)]{tak02} Takeda, Y., 2002, \apj, submitted 
%(astro-ph/0105215)

\bibitem[Th\'evenin(1989)]{the89} Th\'evenin, F., 1989, \aaps, 77, 137

\bibitem[Th\'evenin(1990)]{the90} Th\'evenin, F., 1990, \aaps, 82, 179

\bibitem[Th\'evenin \& Idiart(1999)]{the99} Th\'evenin, F. \& Idiart, T. P.,
1999, \apj, 521, 753

\bibitem[Tukey(1977)]{tuk77} Tukey, J.~W., 1977,
{\it Exploratory Data Analysis}, Addison-Wesley

\bibitem[Ventura \etal(2001)]{ven01} Ventura, P., D'Antona, F., Mazzitelli, 
I. \& Gratton, R., 2001, \apj, 550, 65L

\bibitem[Vogt \etal(1994)]{vog94} Vogt, S.~E. \etal\, 1994, SPIE, 2198, 362

\bibitem[Wallace \etal(1998)]{wal98} Wallace, L., Hinkle, K. \& Livingston, 
W.C., 1998, N.S.O. Technical Report 98-001, 
http://ftp.noao.edu.fts/visatl/README

\bibitem[Weise \etal(1969)]{wei69} Weise, W. L., Smith, M. W., \& Miles, B. M., 
1969, Natl Stand. Ref. Data Ser., Natl Bur. Stand. (U.S.), NSRDS-NBS 22, Vol. II

\bibitem[Weise \etal(1996)]{wei96} Weise, W. L., Fuhr, J. R., \& Deters, T. M.,
1996, J. Phys. Chem. Ref. Data Monograph No. 7 

\bibitem[Weiss \etal(2000)]{wei00} Weiss, A., Denissenkov, P. A., \& 
Charbonnel, C., 2000, \aap, 356, 181

\bibitem[Wolnik \etal(1971)]{wol71} Wolnik, S. J., Berthel, R. O., \&
Wares, G. W., 1971, \apj, 166, L31

\bibitem[Zhao \etal(1998)]{zha98} Zhao, G., Butler, K., \& Gehren, T., 
1998, \aap, 333, 219

\bibitem[Zocalli \etal(1999)]{zoc99} Zocalli, M., Cassisi, G., Piotto, G., 
Bono, G. \& Salaris, M., 1999, \apj, 518, L49

\end{thebibliography}
\end{document}